\pdfoutput=1
\documentclass[%
 reprint,
superscriptaddress,
 amsmath,amssymb,
 aps,
prl,
floatfix,
]{revtex4-2}
\usepackage{graphicx}
\usepackage{dcolumn}
\usepackage{bm}
\usepackage{amsthm}
\usepackage{commath}
\usepackage{mathtools}
\usepackage{graphicx}
\usepackage[colorlinks=true, hyperindex, breaklinks, linkcolor=blue, urlcolor=blue, citecolor=blue]{hyperref} 
\usepackage{amsmath,amsfonts,amssymb,color,graphicx,xcolor,physics}
\usepackage{tikz}
\usepackage{float}
\makeatletter
\let\newfloat\newfloat@ltx
\makeatother
\usepackage{algorithm} 
\usepackage{algorithmicx}
\usepackage[noend]{algpseudocode} 

\newtheorem*{theorem*}{Theorem}
\newtheorem*{corollary*}{Corollary}

\newcommand\nonpfrate[1]{\gamma_{X, Y}}
\makeatletter
\newcommand*{\rom}[1]{\expandafter\@slowromancap\romannumeral #1@}
\makeatother

\let\oldproof\proof
\renewcommand{\proof}{\oldproof}

\def\algbackskip{\hskip-\ALG@thistlm}

\begin{document}


\title{Timescales, Squeezing and Heisenberg Scalings in Many-Body Continuous Sensing}

\author{Gideon Lee}
\affiliation{Pritzker School of Molecular Engineering, The University of Chicago, Chicago, Illinois 60637, USA}

\author{Ron Belyansky}
\affiliation{Pritzker School of Molecular Engineering, The University of Chicago, Chicago, Illinois 60637, USA}

\author{Liang Jiang}
\affiliation{Pritzker School of Molecular Engineering, The University of Chicago, Chicago, Illinois 60637, USA}

\author{Aashish A. Clerk}
\affiliation{Pritzker School of Molecular Engineering, The University of Chicago, Chicago, Illinois 60637, USA}

\date{\today}
\begin{abstract}
The continuous monitoring of driven-dissipative systems offers new avenues for quantum advantage in metrology. This approach mixes temporal and spatial correlations in a manner distinct from traditional metrology, leading to ambiguities in how one identifies Heisenberg scalings (e.g.~standard asymptotic metrics like the sensitivity are not bounded by system size).  Here, we propose a new metric for continuous sensing, the \textit{optimized finite-time environmental quantum Fisher information (QFI)}, that remedies the above issues by simultaneously treating time and system size as finite resources. In addition to having direct experimental relevance, this quantity is rigorously bounded by both system size and integration time, allowing for a precise formulation of Heisenberg scaling.  We also introduce two many-body continuous sensors: the high-temperature superradiant sensor, and the dissipative spin squeezer. Both exhibit Heisenberg scaling of a collective magnetic field for multiple directions.  The spin squeezed sensor has a striking advantage over previously studied many-body continuous sensors: the optimal measurement achieving the full QFI {\it does not} require the construction of a complex decoder system, but can be achieved using direct photodetection of the cavity output field.  

\end{abstract}
\maketitle


\textit{Introduction -- } Quantum metrology \cite{Giovannetti_PRL_2006_qmetrology, Giovannetti_science_2004_qenhanced, Giovannetti_science_2004_qenhanced} is a key application of near-term quantum technologies. In the simplest setting, one prepares an $N$-qubit sensor in an initial state, and then encodes in it an unknown parameter (e.g. by unitary evolution).  The state is then measured, and the results used to estimate the parameter.  Optimizing over measurements and estimators, the minimal estimation error is determined by the quantum Fisher information (QFI) of the state. A fundamental result (the quantum Cramer-Rao bound \cite{Braunstein_PRL_1994_CRB, Paris_intjqi_2009_review, Helstrom_1969_q_estimation_theory}) is that if one assumes the parameter dependence is generated by a sum of single qubit operators, then for an initial product state, the QFI can grow as best as $O(N)$ with increasing $N$ (``standard quantum limit" scaling), whereas for optimally-entangled states, the scaling can be $O(N^2)$ (``Heisenberg limited scaling").  A fundamental challenge in quantum metrology is to find practical states and schemes which can achieve Heisenberg-limited scaling.  

\begin{figure}[ht!]
\centering 
\includegraphics[width=0.95\columnwidth]{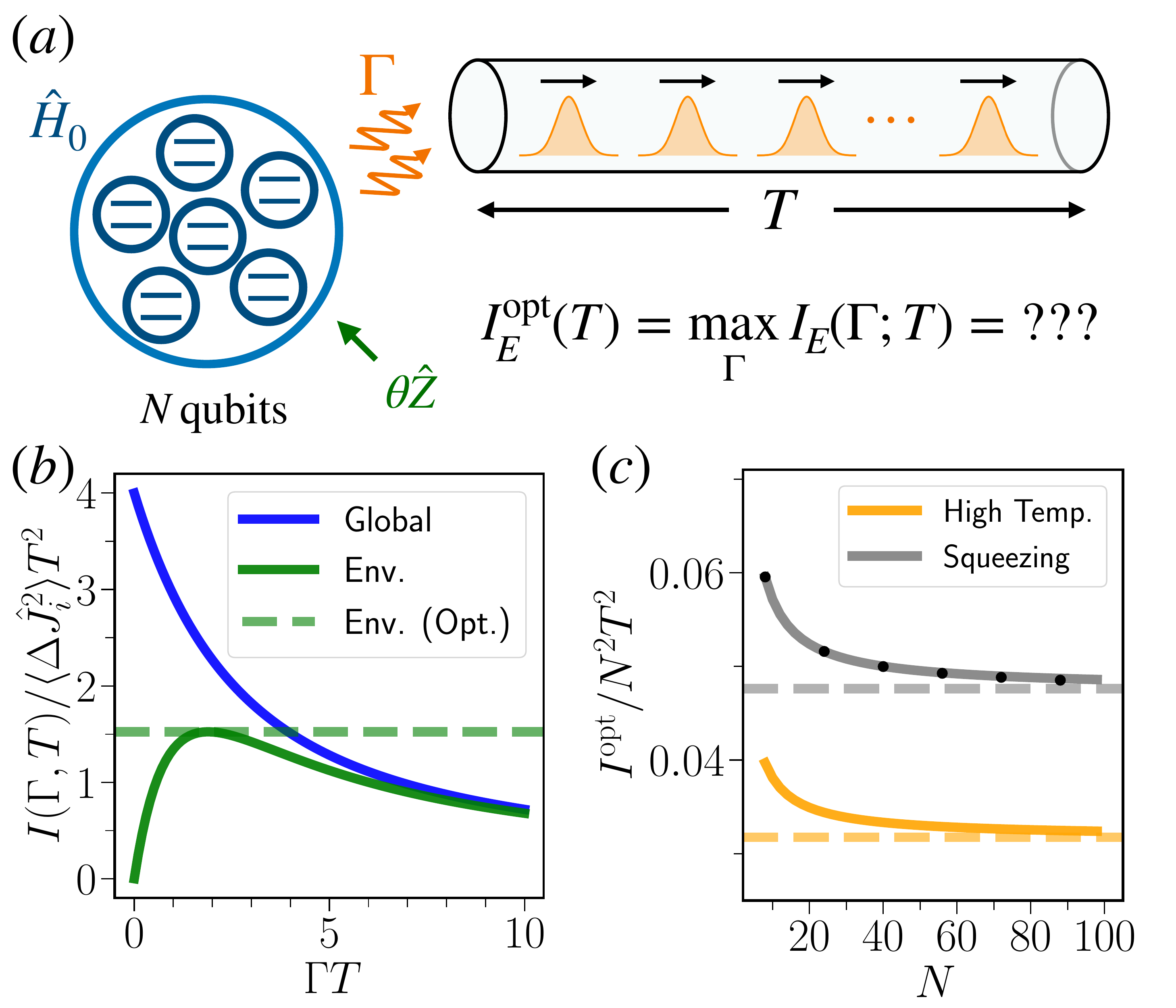} \caption{ \textbf{(a)} Schematic of an $N$-qubit continuous metrology sensor.  At time $T$, the relevant quantum state comprises the qubits as well as wavepackets in the output waveguide.  
\textbf{(b)} Green: environmental QFI $I_E$ (c.f.~Eq.~(\ref{eq:I_E_inf})) versus $\Gamma $, for detecting a global magnetic field with the high-temperature superradiant sensor. $\Gamma$ is the qubit-waveguide coupling. Blue: global QFI $I_G$.    For fixed $T$, $I_E$ has a maximum at $\Gamma \sim 1.92 / T$.  Scaled curves are the same for the spin-squeezed sensor.  
$\langle \Delta \hat{J}_i^2 \rangle$ is the steady state variance of a collective spin component.
\textbf{(c)} Optimized finite-time environmental QFI, $I^{\rm opt}$, versus $N$, for both models in the main text, demonstrating $\propto N^2$ Heisenberg scaling. Solid curves are Eq.~(\ref{eq:opt_model_1}), which is exact for the high-temperature sensor.  For the spin squeezed sensor, we also plot results from full numerical simulations (black dots, $r = \ln (8N)$).
}\label{fig:fig_1} 
\end{figure}

Interest has recently grown in quantum metrological protocols that go beyond the standard ``prepare-and-measure" paradigm, instead performing parameter estimation using continuously monitored systems (see e.g.~\cite{kiilerich_PRA_2014_atomic_photocounting, Catana_IOP_2015_local_asymptotic_normality, Ilias_PRXQ_2022_criticality_enhanced_cts_sensing, Yang_PRX_2023_CQA_sensing, Godley_Quantum_2023_adaptive_measurement, Kiilerich_PRA_2016_bayesian_homodyne, Smiga_arxiv_2024_role_correlations, Smiga_PRR_2023_stochastic_metrology, Radaelli_arxiv_2024_metrology_jump, Radaelli_NJP_2023_correlated_stochastic_FI, Macieszczak_PRA_2016_dynamical_transition_resource, Fernandez_Lorenzo_PRA_2017_close_dissipative_transition, Heugel_PRL_2019_transducer, Garbe_PRL_2020_critical_metrology, Di_Candia_npj_2023_critical_parametric, Yang_arxiv_2025_noisy_CRB, cabot_arxiv_2025_noisy_BTC, Khan_arxiv_2025_tensornetwork_noisy}). 
In this setting (see Fig.~\ref{fig:fig_1}a), the sensor is described by a Hamiltonian that depends on the parameter of interest, and is also coupled to one or more Markovian dissipative environments (which can be concretely viewed as waveguides) \cite{Gammelmark_PRA_2013_cts_meas_metrology, Tsang_PRL_2012_cts_hypothesis_testing}.  The goal is now to estimate the unknown parameter using the information transmitted to the environmental degrees of freedom, i.e.~the photons emitted by the sensor into the waveguides over a time interval $T$.
This setting is extremely natural in numerous experimental platforms, e.g. in cavity QED setups, where one could measure output fields to infer parameters governing the intra-cavity dynamics.  
 Moreover, compared to the the prepare-and-measure approach, continuous sensing may be better suited to tasks that involve complex temporal structures, such as waveform estimation \cite{Tsang_PRL_2011_fund_limit_waveform, Gardner_PRL_2024_waveform_1, Gardner_arxiv_2024_waveform_2}.

There are by now well-established methods for calculating the QFI relevant to continuous metrology protocols, and in some cases, even corresponding optimal measurement protocols \cite{Gammelmark_PRA_2013_cts_meas_metrology, Gammelmark_PRL_2014_cts_meas_QFI, Yang_PRX_2023_CQA_sensing}.  The key remaining challenge would thus seem to be to identify many-body sensors that achieve optimal QFI scalings that, crucially, are also attainable using experimentally-tractable measurements.

In this paper, we address this issue, introducing several multi-qubit setups that allow optimal continuous metrology using simple measurements.  We also address an even more basic question: 
{\it In continuous metrology, what exactly constitutes a ``good" QFI, and what is a useful definition of  Heisenberg scaling}?  We argue that similar to standard metrology, simply finding a QFI scaling of $N^2$ or higher is not particularly meaningful if resources are not properly constrained.  Previous works have focused on the long-time growth rate of the QFI, the so-called sensitivity.  This quantity can trivially exhibit super-Heisenberg scaling in regimes that have no practical utility, motivating the search for better metrics.  

We propose an alternative figure of merit for continuous sensing, the optimized finite-time environmental QFI.  It cannot be trivially made to have arbitrary scalings with $N$, and also has the advantage of being directly relevant to experiment, as it explicitly incorporates bandwidth constraints (i.e.~it characterizes optimal performance given a certain fixed sensing time).  

\textit{Preliminaries-- } Continuous metrology aims to perform parameter estimation by coupling a sensor (taken here to be $N$ qubits) to a dissipative environment whose state can be monitored; without loss of generality, we may regard these baths as photonic waveguides.  We focus on the standard setting where the goal is to estimate a single infinitesimal Hamiltonian parameter.  The Hamiltonian of the isolated sensor will have the form
\begin{equation}\label{eq:boring_H_def}
    \hat{H} = \hat{H}(\theta) = \hat{H}_0 + \theta \hat{Z},
\end{equation}
where $\theta$ is the parameter of interest.  
We refer to $\hat{Z}$ as the generator of the parameter $\theta$, and take it to be dimensionless (so that $\theta$ has units of frequency). 

We next couple the sensor to measurement waveguides, which to the sensor qubits look like Markovian dissipation.  The sensor couples to each of these waveguides via an operator $\hat{L}_i$, which represent the sensor quantity monitored by each waveguide.  The evolution of the sensor qubits alone (i.e. tracing out the waveguides) is then given by the GKSL (Lindblad) master equation
\begin{equation}
    \label{eq:Lindblad}
    \partial_t \hat{\rho} = - i [\hat{H}, \hat{\rho}] +  \sum_i \Gamma_i 
    \mathcal{D}[\hat{L}_i] \hat{\rho},
\end{equation}
where $\hat{D}[\hat{L}](\cdot) = \hat{L}(\cdot)\hat{L}^{\dag} - \{\hat{L}^{\dag} \hat{L}, \cdot\}/2$. We take the $\hat{L}_i$ to be dimensionless, and have introduced coupling constants $\Gamma_i$ with units of rate.  
These determine the qubit-waveguide coupling strengths, and the rate at which each quantity $\hat{L}_i$ is effectively monitored.  We will view these as experimentally-controllable parameters to be optimized, and thus the normalization of each jump operator will not play a role in what follows.  

With these definitions, we can now quantify our ability to estimate $\theta$ after some time $T$ using the QFI
associated with the state of the sensor and the waveguide (working under the standard assumption that we have access to the full state of each waveguide).  Ref.~\cite{Gammelmark_PRL_2014_cts_meas_QFI} showed that this could be achieved by only looking at the dynamics of the sensor qubits, via 
a pseudo-density matrix $\hat{\mu}(t)$ obtained from the
so-called two-sided master equation \footnote{Note however, that Ref.~\cite{Gammelmark_PRL_2014_cts_meas_QFI} considered a more general situation where the Hamiltonian and jump operator may also be time-dependent}: 
\begin{equation}\label{eq:cts_2sided}
      \partial_t \hat{\mu} \equiv -i \hat{H}_0 \hat{\mu} + i \hat{\mu} \hat{H}(\theta)
  +  \sum_i  \Gamma_i \mathcal{D}[\hat{L}_i] \hat{\mu}. 
\end{equation} 
While this equation is not a valid CPTP master equation, it can be given a simple physical interpretation: it describes the dephasing of an auxiliary qubit $\sigma$ that couples to the sensor via an interaction $\hat{H}_{\rm int} = (\theta/2) \hat{Z} \hat{\sigma}_z$
(see SM \cite{SM}).
Throughout this work, we focus on the case where the sensor qubits start in the (unique) dissipative steady state of Eq.~(\ref{eq:Lindblad}) at $\theta = 0$.  This corresponds to a quench protocol.  For $t < 0$, $\hat{H} = \hat{H}(0)$, and the system reaches its dissipative steady state.  Then at $t=0$ the Hamiltonian is suddenly switched to $\hat{H} = \hat{H}(\theta)$, and the sensing protocol begins. 
WLOG, we will always shift $\hat{Z}$ so it has zero mean in the $\theta=0$ steady state.

One can use $\hat \mu(t)$ to now calculate the two QFIs relevant to continuous metrology
\footnote{The initial condition corresponds to the initial state of the sensor qubits.  We make the natural assumption that this is the steady state of Eq.~XXX, i.e. before $\hat{H}(\theta)$ is turned on, the sensor is in the dissipative steady state determined by $\hat{H}_0$ and the coupling to the waveguides.}
.  The first is the global QFI $I_{\rm G}$, which involves the full quantum state of the sensor qubits and the waveguides.  The second is perhaps more experimentally relevant: this is the ``environmental" QFI $I_{\rm E}$, and is based on {\it only} allowing measurement of the waveguides. Ref.~{\cite{Yang_PRX_2023_CQA_sensing}} showed this can also be obtained from $\hat \mu(t)$. 
The general expression for these QFIs is $(\alpha = {\rm E},{\rm G})$, 
\begin{equation}\label{eq:cts_QFI_expression}
  I_{\alpha}[\hat{Z}](\{\Gamma_i\}; T) = -4 \partial_{\theta}^2 \mathcal{F}_{\alpha}(\theta, T) |_{\theta = 0}, 
\end{equation}
with
\begin{equation}\label{eq:F_G}
  \mathcal{F}_{\rm G}(\theta, T) =  | {\rm Tr} \hat{\mu}_{\theta} (T) |,
\end{equation}
\begin{equation}\label{eq:F_E}
  \mathcal{F}_{\rm E}(\theta, T) = {\rm Tr} \left( \sqrt{\hat{\mu}_{\theta}(T) \hat{\mu}^{\dag}_{\theta}(T)} \right).
\end{equation}
where the subscript $\theta$ indicates that $\hat{\mu}_{\theta}$ was computed by evolving the two-sided master equation with parameter $\theta$. 


\textit{Heisenberg scaling in continuous metrology-- }
We now ask the crucial question of how best to quantify the performance of a continuous sensor, and to identify Heisenberg scaling in the large-$N$ limit.  
To meaningfully discuss scaling we must  constrain how $\hat{Z}$ changes with increasing qubit number. Similar to standard quantum metrology, we take it to be a permutation-symmetric sum of single-qubit operators, i.e.~a collective angular momentum operator:
$\hat{Z} = \frac{1}{2} \sum_{j=1}^N \vec{r} \cdot \vec{\hat{\sigma}}^{(j)} \equiv \hat{J}_{\vec{r}}$,
where $\vec{\hat{\sigma}}^{(j)} $ is the vector of Pauli operators for qubit $j$, and $\vec{r}$ is some unit vector. 
In standard metrology, it is common to consider performance in either the large-$N$ or large-$T$ limit (where $T$ controls the how the parameter dependence is imprinted on the sensor state).  In continuous metrology, $T$ plays a different role, as it also controls the ``size" of the final sensor state (which describes the $N$ qubits as well as the photons emitted to the waveguide).  Previous work~\cite{Yang_PRX_2023_CQA_sensing, Godley_Quantum_2023_adaptive_measurement} has characterized continuous sensors by focusing on the asymptotic large-$T$ limit, where $I_{\rm E/G}(T) \rightarrow \mathcal{S}_Z T$, with the constant $\mathcal{S}_Z$ (the ``sensitivity") determined by the steady-state noise spectrum of $\hat{Z}$ at zero frequency: $\mathcal{S}_Z = 2 S_{ZZ}[\omega = 0]$ \cite{SM}.  It is tempting to also take the large-$N$ limit, and then define Heisenberg versus SQL scaling by whether  $\mathcal{S}_Z \sim O(N)$ or 
$\sim O(N^2)$, as has has been done in many works (see e.g.~\cite{Yang_PRX_2023_CQA_sensing, Cabot_PRL_2024_BTC_sensing, cabot_arxiv_2025_noisy_BTC}).  

While the sensitivity metric might seem natural, it has several undesirable features.  First, even with the constraints we have placed on the generator $\hat{Z}$, there is no fundamental constraint on how quickly $\mathcal{S}_Z$ can scale with $N$. 
As 
$\mathcal{S}_Z$ is proportional to a noise spectrum, it has units of time.  It can thus exhibit arbitrarily fast $N$ scalings simply by having the characteristic timescales of the system diverge with qubit number.  While $\langle \hat{Z}^2 \rangle$ cannot grow faster than $N^2$, there is no corresponding bound on $\mathcal{S}_Z$.
This is also problematic at a practical level:  the utility of the asymptotic large-$N$, large-$T$ limit is questionable if it is only obtained for $T \gg O(N^\alpha)$ for $\alpha >0$.  Further, this dependence on an effective timescale also makes the $\mathcal{S}_Z$ crucially dependent on how exactly one chooses to scale the couplings $\Gamma_j$ with $N$ (see \cite{SM} for concrete examples of how different choices here lead to arbitrary $N$ scalings of $\mathcal{S}_Z$).  In general $\mathcal{S}_Z$ can be arbitrarily large for a very slow sensor, a situation which is of little practical interest.  

Given these pathologies with using the sensitivity to characterize continuous sensors, we propose a different metric.  Instead of focusing on the large $T$ limit, we fix the measurement time $T$ to an $N$-independent value, and ask what the best possible {\it environmental} QFI is at this time, {\it optimizing over the choice of waveguide couplings $\Gamma_i$}.  These choices directly connect to experimental utility.  Using a fixed $T$ means that we will not erroneously conclude that an extremely slow sensor is optimal; it also lets one incorporate bandwidth constraints relevant to the particular sensing target.  Focusing on the environmental QFI also matches the philosophy of continuous sensing (i.e.~the waveguides represent measurement channels) and also in many cases simplifies the experiment (e.g.~in cavity QED, it is natural to probe output photons using photodetection, heterodyne or homodyne measurements, but less easy to directly measure intracavity degrees of freedom).  Finally, optimizing over the choice of $\Gamma_i$ is consistent with the general philosophy that QFI involves optimizing over measurements: we view the waveguides and the couplings to them as part of the generalized measurement of the system qubits.  This optimization also means that we are not sensitive to how one normalizes the jump operators.

We thus will characterize $N$-qubit continuous sensors by their optimized finite-time environmental QFI $I_{\rm E}^{\rm opt}(T)$:
\begin{equation}
    I_{\rm E}^{\rm opt}(T) \equiv \max_{ \{\Gamma_j\} } \, I_{\rm E} (\{\Gamma_i\}; T), 
\end{equation}
Note that when optimizing over the $\Gamma_j$, there will be a non-trivial maximum representing a tradeoff between two competing effects.  For $\Gamma_j \rightarrow 0$, there is no coupling to the waveguides, and $I_{\rm E}$ clearly vanishes.  However, for very large $\Gamma_j$ the induced dissipation on the system qubits will hinder the ability of $\hat{H}(\theta)$ to generate a parameter dependence, and $I_{\rm E}$ will also generically vanish.  One might crudely guess an  optimal choice of $\Gamma_j \sim 1/T$, something that we show to be true in several concrete examples.  Note this means that $I_{\rm E}^{\rm opt}(T)$ will not coincide with the sensitivity, as the optimal regime is not one where $T$ is larger than all other timescales.   

We can now meaningfully ask about sensor performance in the large-$N$ limit by asking how  $I_{\rm E}^{\rm opt}(T)$ scales with $N$ for fixed $T$.  Using the fact that $I_E(T) \leq I_G(T)$ and constraints on correlation functions, one can prove the (loose) bound \cite{SM}:
\begin{equation}\label{eq:loose_bound}
    I_{\rm E}^{\rm opt}(T) \leq N^2 T^2,
\end{equation}
where the prefactor of unity comes from fixing $\hat{Z} = \hat{J}_{\vec{r}}$.
We stress that there is no such generic bound on the sensitivity $\mathcal{S}_Z$.  This motivates defining Heisenberg-limited scaling with $N$ for our continuous sensor as a sensor that exhibits $I_{\rm E}^{\rm opt} \propto N^2$ in the large $N$ limit.  Note that on dimensional grounds, one can argue that generically
$I_{\rm E}^{\rm opt}(T) \sim \langle  \hat{Z}^2   \rangle T^2$ \cite{SM}.  This suggests a very simple prerequisite for optimal sensor design: the dissipative steady state of the sensor should yield a maximum variance of $\hat{Z}$, scaling as $\propto N^2$. However, we note that this is not a sufficient condition \footnote{For example, the sensor proposed in \cite{Cabot_PRL_2024_BTC_sensing} has a steady state which has $N^2$ variance in $\hat{J}_{y, z}$, but does not exhibit a Heisenberg scaling along those axes.}.

\begin{figure}[t!]
\centering \includegraphics[width=0.9\columnwidth]{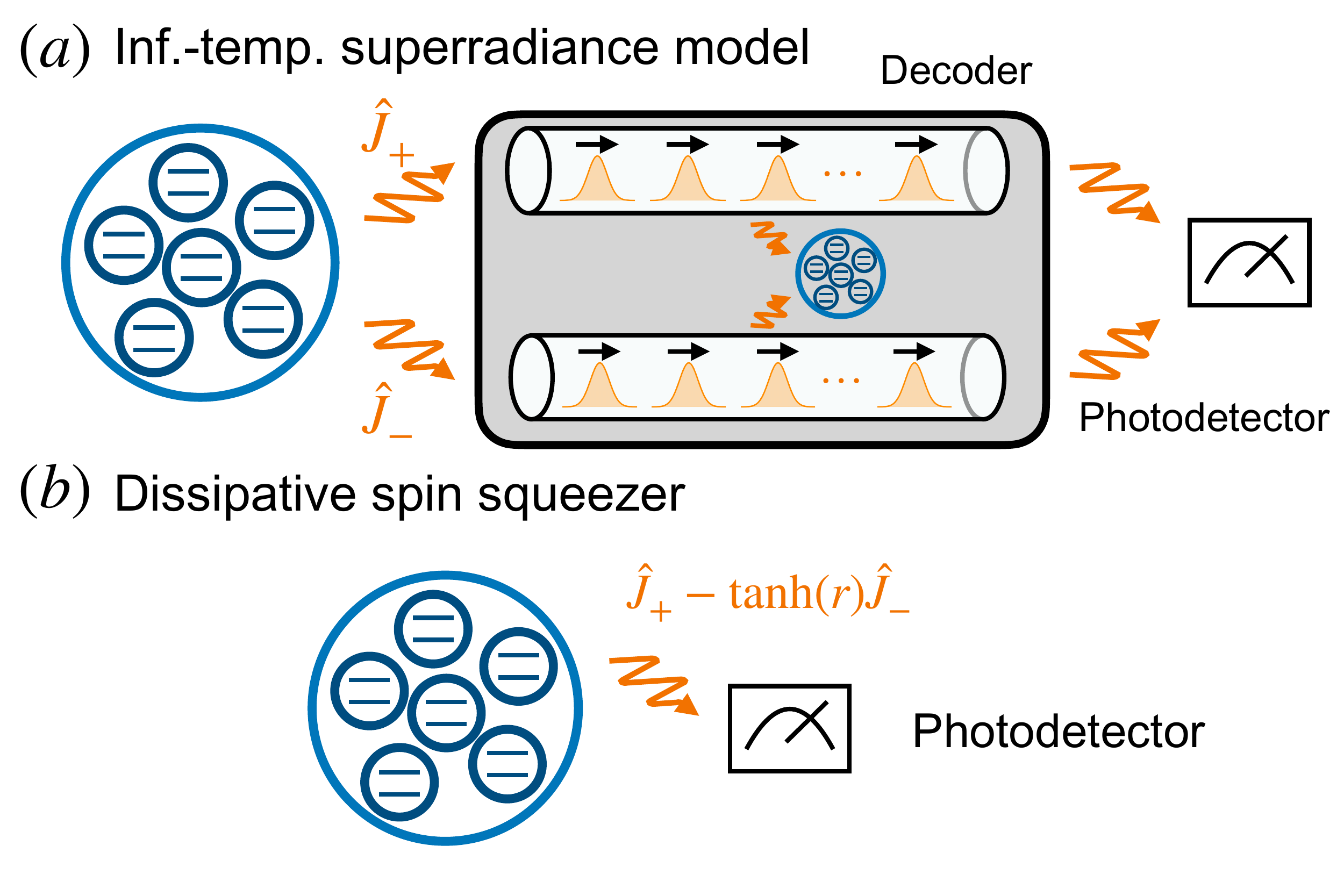} \caption{ \textbf{(a)} Schematic depicting the complete-dephasing sensor. The optimal measurement involves engineering the decoder via the recipes in \cite{Yang_PRX_2023_CQA_sensing, Stannigel_NJP_2012_CQA_OG, Roberts_PRXQ_2021_hTRS}, which involves fine-tuning the coupling to two independent non-reciprocal waveguides and a downstream system. Photodetection is then performed on the output field of these waveguides. \textbf{(b)} Schematic depicting the dissipative spin squeezer. The optimal measurement simply involves performing photodetection directly on the output field of the sensor.}
\label{fig:fig_2} 
\end{figure}

\textit{Calculating the optimized environmental QFI-- }
We now turn to calculating $I_{\rm E}^{\rm opt}(T)$ for several examples of many-body qubit sensors.  Unfortunately, while $I_{\rm E}$ is more experimentally-relevant than the global QFI $I_{\rm G}$, $I_G$ is far easier to calculate, as it can be obtained directly from the sensor's symmetrized stationary autocorrelation function $C_{ZZ}(t) = \langle \{ \hat{Z}(t), \hat{Z}(0) \}\rangle$ \cite{SM, Gammelmark_PRA_2013_cts_meas_metrology}: 
\begin{equation}\label{eq:GlobalQFINoise}
\begin{aligned}
    I_{G}(T) = &4 \int_0^T d \tau_1 \int_0^{\tau_1} d \tau_2 C_{ZZ}(\tau_2) 
\end{aligned}
\end{equation}
This is simply the variance of $\int_0^T dt' \hat{Z}(t')$.  In general, a similarly simple expression for $I_{\rm E}$ does not exist:
unlike $I_{\rm G}$, it is not linear in the sensor density matrix (c.f.~Eqs.~(\ref{eq:F_G}),(\ref{eq:F_E})).  However, we find that simple expressions can be found for a restricted class of relevant systems.    These are sensors where $\hat{H}_0=0$ (i.e.~the only Hamiltonian is the $\theta$-dependent perturbation), and where the dissipative steady state is either fully mixed, or completely pure.  In the fully mixed case, we also require that the $\hat{L}_i$ operators can be chosen to be Hermitian.  For such systems, 
we find \cite{SM}:
\begin{equation}
\begin{aligned}
    I_{E}(T) = 
    I_{G}(T)     
    - 4 \int_0^T d \tau_1 \int_0^{\tau_1} d \tau_2 C_{ZZ}(\tau_1 + \tau_2),
\end{aligned}
\label{eq:IESimple}
\end{equation}
We now discuss two many-body systems relevant to cavity QED experiments where this formula applies, and where our sensor exhibits Heisenberg-like scaling 
$I_{\rm E}^{\rm opt}(T) \propto N^2$. Note that by rewriting Eq.~(\ref{eq:IESimple}) in terms of a filter function acting on the noise spectral density, one may tighten Eq.~(\ref{eq:loose_bound}) to $I_E(T) \lesssim 0.262 N^2 T^2$ \cite{SM}.

\textit{High-temperature superradiant sensor -- } In what follows, we consider collective $N$-qubit systems, where Eq.~(\ref{eq:Lindblad}) can be fully expressed in terms of collective angular momentum operators 
$\hat{J}_\alpha$, and 
restrict attention to the subspace of maximum total angular momentum. It was recently shown that a model of this form describing collective Rabi driving balanced by collective (superradiant) loss could be a good continuous sensor \cite{Cabot_PRL_2024_BTC_sensing, Iemini_PRA_2024_time_crystal_sensor}. This model exhibits
non-trivial phase transition physics and even time-crystal like behavior in the large $N$ limit \cite{Iemini_PRL_2018_OG_timecrystal}.  One might wonder if those features are relevant to its performance as a sensor.  

To test this, we consider an even simpler model, where there is no coherent Rabi driving of the spins, but instead collective incoherent pumping.  We also constrain the model so that the steady state is maximally mixed (something we already anticipated should be optimal)  The sensor's master equation has the form:
\begin{equation}
\label{eq:SuperradianceMEQ}
\begin{aligned}
    \partial_t \hat{\rho} = 
    -i \theta [ \hat{Z}, \hat{\rho}] + 
    \Gamma \mathcal{D}[\hat{J}_{-}] \hat{\rho} + \Gamma \mathcal{D}[\hat{J}_{+}] \hat{\rho}.
\end{aligned}
\end{equation}
where $\hat{J}_\pm = \hat{J_x} \pm i \hat{J}_y$.   For $\theta = 0$, this corresponds to a high-temperature superradiant decay model.  In terms of the sensing setup, it implies that there are two waveguides coupled to the system (one with a Jaynes-Cumming (JC) coupling, the other with an anti-JC coupling).  

It is possible to re-write the master equation using only Hermitian jump operators, hence it satisfies all the conditions for using Eq.~(\ref{eq:IESimple}) to calculate $I_E(T)$.  We consider sensing a weak magnetic field in either the $x,y$ or $z$ directions, corresponding to $\hat{Z} = \hat{J}_{x,y,z}$.  While the model is not linear, one can nonetheless exactly calculate the relevant stationary autocorrelation functions:
\begin{equation}
        C_{J_{\alpha} J_{\alpha}}(\tau)
        = 2 \langle \Delta \hat{J}_{\alpha}^2 \rangle e^{- \Gamma_{\alpha} \tau},
\end{equation}
where the steady state variances in the infinite-temperature steady state are $\langle \Delta \hat{J}_{x, y, z}^2 \rangle = J(J+1)/3 \simeq N^2/12$ and $\Gamma_z = 2 \Gamma, \Gamma_{x, y} = \Gamma$. Using Eq.~(\ref{eq:IESimple}), we find:
\begin{equation}\label{eq:I_E_inf}
\begin{aligned}
    &I_{\rm E}[\hat{J}_{\alpha}](\Gamma; T) = \\
    &4 T^2
    \langle \Delta \hat{J}_{\alpha}^2 \rangle    
    \left[
        \frac{2}{\Gamma_{\alpha} T}  - 
        \frac{1}{ \left( \Gamma_{\alpha} T \right)^2}
        \left( 
            3 - 4 e^{- \Gamma_{\alpha} T} +  e^{- 2 \Gamma_{\alpha} T}  \right) \right].
\end{aligned}
\end{equation}
For fixed $N$, $I_E / T^2$ is just a function of the dimensionless parameter $\Gamma_{{\alpha}} T$.  
It is straightforward to check that for fixed $N,T$, $I_E$ vanishes as expected as $\Gamma_\alpha \rightarrow 0$ ($I_E \propto \Gamma_\alpha^2$)  or $\Gamma_\alpha \rightarrow \infty$ ($I_E \propto 1/\Gamma_\alpha$).  $I_E(T)$ is maximized at 
$\Gamma_{\alpha} \simeq 1.89 / T$, yielding an optimized environmental QFI at time $T$ of:
\begin{equation}\label{eq:opt_model_1}
    \begin{aligned}
        I^{\rm opt}_{\rm E}[\hat{J}_{x, y, z}](T) \simeq 0.191 \frac{2J(J+1)}{3} T^2 \sim O(N^2 T^2).
    \end{aligned}
\end{equation}
We thus obtain Heisenberg scaling in the number of qubits,  $I_E \propto N^2 T^2$, for all three magnetic field directions.  Note this is in contrast to the model studied in Ref.~\cite{Cabot_PRL_2024_BTC_sensing}, where Heisenberg-like scaling is only found for one direction of the field to be sensed. Additionally, Ref.~\cite{Cabot_PRL_2024_BTC_sensing} attributes their Heisenberg scaling to persistent oscillations in the time-crystal phase; here, we find in a similarly collective model that such persistent oscillations are not necessary for attaining Heisenberg scaling.

In Fig.~(\ref{fig:fig_1}b), we plot both the environmental QFI $I_{\rm E}(T)$ and the global QFI $ I_{\rm G}(T)$ as a function of $\Gamma_\alpha$ for fixed $T$.  The global QFI does not vanish for small $\Gamma$, but in this limit the information resides completely in the sensor qubits, and is not accessible through measurements of the waveguide.  While we have argued that $I^{\rm opt}_E(T)$ is the most meaningful metric for our sensor, it is also interesting to consider sensitivity.
It follows from Eq.~(\ref{eq:GlobalQFINoise}) that $\mathcal{S}_{J_\alpha} = 8 \langle \Delta J_\alpha^2 \rangle / \Gamma_\alpha$.  The sensitivity can be made arbitrarily large by making $\Gamma_\alpha \rightarrow 0$.  As discussed, this is a misleading result, as the asymptotic long-time limit characterized by the sensitivity only sets in for time $T \gg 1 / \Gamma_\alpha$.   

We now turn to a crucial question that goes beyond calculating the optimized QFI:  {\it what} exactly one must measure in the waveguides to extract the minimal estimation error?  We are interested in systems where the optimal measurement is both known and experimentally feasible.   As discussed in   Ref.~\cite{Yang_PRX_2023_CQA_sensing}, one can in principle construct an optimal measurement by building a ``coherent quantum absorber" system \cite{Stannigel_NJP_2012_CQA_OG} that absorbs all photons in the waveguide if $\theta = 0$, and then performing photodetection on photons that make it past the absorber.  The issue is that finding the needed absorber system is in general an intractable task.  However, for a class of systems, the absorber system is essentially the {\it same} as the original system up to sign changes in the system Hamiltonian \cite{Stannigel_NJP_2012_CQA_OG,Roberts_PRXQ_2021_hTRS,Tsang_arxiv_2024_quantum_reversal}.  It would a priori seem unclear when and why this trick would work (though several examples are presented in Ref.~\cite{Yang_PRX_2023_CQA_sensing}).

We note here that the construction of a simple absorber system (and hence optimal QFI-saturating measurement) is possible for any system having a so-called ``hidden time-reversal symmetry", a version of quantum detailed balance \cite{Roberts_PRXQ_2021_hTRS}.  Recent work (motivated by finding exact solutions of non-equilibrium steady states) has by now shown this symmetry holds in a number of non-trivial models, ranging from nonlinear driven cavity models \cite{Roberts_PRX_2020_HTRS_blockade}, to many body bosonic \cite{Roberts_PRL_2023_many_boson_hTRS} and spin \cite{Yao_PRL_2025_XXZ_HTRS, Roberts_PRL_2023_inf_range_TFIM, Lingenfelter_PRX_2024_spin_ent} models.  This provides a wealth of systems that could be interesting for continuous metrology.  Returning to our
high-temperature superradiant sensor in Eq.~(\ref{eq:SuperradianceMEQ}), 
it can be easily shown to have hTRS \cite{SM}, and hence allows for an easy construction of the absorber and ideal measurement setup (see Fig.~\ref{fig:fig_2}(a)). 

Finally, note that the high-temperature superradiance model can be extended to a one-parameter family of models consistent where Eq.~(\ref{eq:IESimple}) holds.  These are described by the master equation:
\begin{equation}
\begin{aligned}
    \partial_t \hat{\rho} = - i \theta[\hat{Z}, \hat{\rho}] + 2 (1 + \eta) \Gamma  \mathcal{D}[\hat{J}_{x}] \hat{\rho} + 2 (1 -\eta)\Gamma \mathcal{D}[\hat{J}_{y}] \hat{\rho}.
\end{aligned}
\end{equation}
where the parameter $\eta \in (-1,1)$.  This reduces to Eq.~(\ref{eq:SuperradianceMEQ}) when $\eta = 0$.  These models have an identical $I_E^{\rm opt}$ no matter what the value of $\eta$ \cite{SM}.  

\bigskip

\textit{Dissipative spin squeezer-- } The second collective $N$-qubit sensing model we consider describes dissipative spin squeezing \cite{Groszkowski_PRX_2022_dissipative_spin_squeezing, Agarwal_PRA_1990_OG_diss_ss, Dalla_Torre_PRL_2013_exp_spin_sq_diss}:
\begin{equation}
    \label{eq:SpinSqzMaster}
    \partial_t \hat{\rho} = - i \theta[\hat{Z}, \hat{\rho}] + \Gamma \mathcal{D}[\hat{J}_+ - \tanh(r) \hat{J}_-] \hat{\rho}.
\end{equation}
This mimics driving the qubits with broadband squeezed light with quadrature squeezing $e^{-2r}$, and for even $N$, results in a pure steady state that exhibits strong spin squeezing of $\hat{J}_y$, allowing sub-SQL estimation of a $x$ magnetic field.  For large $r$, the steady state exhibits Heisenberg-limited spin squeezing. We focus on $N$ even \footnote{In contrast, $N$ odd does not yield a pure steady state \cite{Groszkowski_PRX_2022_dissipative_spin_squeezing}. Nevertheless, numerical results suggest that $N$ odd also yields a finite-time environmental QFI that goes like $N^2$, although this is far less useful than the $N$ even case, see \cite{SM} for further comments.}, which allows us to apply Eq.~(\ref{eq:IESimple}).

This dynamics has previously been studied as a means for preparing squeezed states, to be then used in traditional metrological protocols.  Here, we imagine using the same setup for continuous metrology.  Concretely, the dynamics in Eq.~(\ref{eq:SpinSqzMaster}) can be realized by engineering both JC and anti-JC couplings to a single cavity mode that is strongly damped (see e.g. \cite{Dalla_Torre_PRL_2013_exp_spin_sq_diss, Groszkowski_PRX_2022_dissipative_spin_squeezing}); the cavity damping can come from coupling to a waveguide.  Our continuous metrology protocol would then involve monitoring the light emitted from the waveguide coupled to the cavity.  

As the evolution here is purely dissipative and leads to a pure steady state, we can again compute the environmental QFI directly from correlation functions using Eq.~(\ref{eq:IESimple}).  Given that the steady state has a net polarization in the $\hat{J}_z$ direction and is squeezed in $\hat{J}_y$, it will be an optimal sensor for a small magnetic field in the $x$ direction.  We thus consider a generator $\hat{Z} = \hat{J}_x$, and need to calculate the stationary autocorrelation function of this object.  We can do this approximately using a leading-order cumulant expansion \cite{SM}, finding simple exponential decay.  In the large-$r$ limit of interest, we have:
\begin{equation}
    C_{J_x J_x}(\tau ) \simeq 2 \langle \Delta \hat{J}_x^2 \rangle \exp(- 2 \Gamma \tau).
\end{equation}
where for large-$r$, the steady state variance is $\langle \Delta \hat{J}_x^2 \rangle \rightarrow J(J+1)/2 \simeq N^2/8$. 
Given the exponential correlation function, optimization over $\Gamma$ is analogous to the thermal superradiant setup, hence this system similarly exhibits Heisenberg scaling along the $x$-axis,
\begin{equation}\label{eq:opt_model_2}
    I_{\rm E}^{\rm opt}[\hat{J}_x](T) \simeq 0.191 \frac{J(J+1)}{2} T^2 \sim O(N^2 T^2).
\end{equation}
Note the prefactor here is larger that the thermal superradiant sensor by $3/2$, reflecting the larger steady state variance of $\hat{J}_x$ in a strongly spin squeezed state versus the high-temperature state.  
While these results are based on a cumulant approximation, we show in Fig.~\ref{fig:fig_1}b a comparison against a full numerical simulation of Eq.~(\ref{eq:SpinSqzMaster}), which shows an excellent agreement.  

While it might seem that the spin squeezing setup only provides a modest prefactor improvement over the thermal superradiant sensor, there is another more significant advantage: the optimal measurement needed to achieve the QFI scaling is much easier.  As discussed, for the high-temperature model, reaching optimal QFI with photodetection requires creating a second copy of the original system, and coupling them via two cascaded non-reciprocal wavegudes (see. Fig.~\ref{fig:fig_2}b).  In contrast, no doubled-system construction is needed when using the dissipative spin squeezer. Here, one can saturate the QFI by directly performing photodetection on the output field. This result \footnote{This is implicit in the decoder construction of \cite{Yang_PRX_2023_CQA_sensing}; we provide a complementary proof in \cite{SM}} holds for any system that hosts a pure dark state, and hence holds for $N$ even. Finally, we note that in the large $r$ limit, we may obtain similar results if one instead tried to sense a $z$ magnetic field, i.e.~for the choice  
$\hat{Z} = \hat{J}_z$: one has Heisenberg-limited scaling, and the optimal measurement is again simple photodetection.  

\textit{Discussion -- } In this work, we re-examined the question of Heisenberg scalings in continuous metrology, stressing the importance of timescales.  We proposed a figure of merit, $I_{\rm opt}^{\rm E}(T)$, that removes ambiguity by providing a proper accounting of time as a finite resource. $I_{\rm opt}^{\rm E}(T)$ cannot be trivially made to have arbitrary scalings with $N$, and naturally incorporates bandwidth constraints. One striking finding is that the proper accounting of resources leads to an optimal environmental QFI that scales with $T^2$ for finite times, in contrast with the asymptotic $T$ scaling. We then proposed two simple many-body sensors that exhibit Heisenberg scaling in the continuous setting.

What can we learn from the two sensors we analyzed that might generalize and help in the search for additional many-body continuous sensors? First, for a large QFI, one ought to look for sensors that host a steady state with high variance. This is similar to traditional metrology, but the great boon of continuous sensing is that \textit{the steady state can be mixed}. The second takeaway message is that pure dark states are a great advantage in continuous metrology, since such systems can achieve the QFI bound by simple photodetection measurements, and without the use of complex decoder setups.  An important question we leave to future work is the calculation of $I^{\rm opt}_{\rm E}(T)$ for these models and otherwise in the presence of noise. Finally, while derived in a fairly restricted setting, Eq.~(\ref{eq:IESimple}) hints at a more general relationship between between $I_{\rm E}$ and the sensor's noise spectrum \cite{SM}; we leave the exploration of this curious connection for future work.

\textit{Acknowledgements -- } We thank Andrew Pocklington, Guo Zheng, Anjun Chu, and Mikhail Mamaev for useful discussions. 
This work was primarily supported by the DOE Q-NEXT Center (Grant No. DOE 1F-60579). A.~C. acknowledges support from the Simons Foundation (Grant No. 669487, A. C.).
L.J. acknowledges support from the ARO (W911NF-23-1-0077), ARO MURI (W911NF-21-1-0325), AFOSR MURI (FA9550-21-1-0209, FA9550-23-1-0338), NSF (ERC-1941583, OMA-2137642, OSI-2326767, CCF-2312755, OSI-2426975), and Packard Foundation (2020-71479).

\bibliographystyle{apsrev4-2}
\bibliography{references}


\clearpage

\widetext
\begin{center}
\textbf{\large Supplemental Material: ``Timescales and Heisenberg Scalings in Many-Body Continuous Sensing''}
\end{center}
\setcounter{equation}{0}
\setcounter{figure}{0}
\setcounter{table}{0}
\setcounter{page}{1}
\makeatletter
\renewcommand{\theequation}{S\arabic{equation}}
\renewcommand{\thefigure}{S\arabic{figure}}
\renewcommand{\bibnumfmt}[1]{[S#1]}
\renewcommand{\citenumfont}[1]{S#1}


\section{Metrology as the dual of noise}

\subsection{Continuous sensing and qubit dephasing}

In this appendix, we perform the mapping of the two-sided master equation to a qubit dephasing problem, in order to obtain the long-time sensitivity in terms of the noise spectrum. We will explicitly focus on obtaining the global QFI; however, since the sensitivity is a long-time expression, this also applies to the environmental QFI. We begin with the two-sided master equation for the Hamiltonian sensing problem, 
\begin{equation}\label{eq:H_2sided_1}
  \begin{aligned}
      \partial_t \hat{\mu} =  -i \left[ \hat{H}(\theta_1) \hat{\mu} - \hat{\mu} \hat{H}(\theta_2) \right] + \sum_{j} \mathcal{D}[\hat{L}_j] \hat{\mu},
  \end{aligned}
\end{equation} 
where $\hat{H}(\theta) = \hat{H}_0 + \theta \hat{Z}$. Note that for the two-sided master equation presented in the main text, we set $\theta_1 = 0$. Defining \begin{equation}
  \begin{aligned}
    \theta_s &= \frac{1}{2} (\theta_1 + \theta_2), \\
    \theta_d &= \frac{1}{2} (\theta_1 - \theta_2),
  \end{aligned}
\end{equation}
we can rewrite Eq.~(\ref{eq:H_2sided_1}) as 
\begin{equation}\label{eq:H_2sided_2}
  \begin{aligned}
      \partial_t \hat{\mu} =  -i \left[ \hat{H}_0 + \theta_s \hat{Z}, \hat{\mu} \right] - i \theta_d \left( \hat{Z} \hat{\mu} + \hat{\mu} \hat{Z} \right) + \sum_{j} \mathcal{D}[\hat{L}_j] \hat{\mu}.
  \end{aligned}
\end{equation} 
The key observation is the following: while Eq.~(\ref{eq:H_2sided_2}) is not a physical master equation, it can be obtained from a physical master equation by introducing an ancilliary qubit. 

In particular, let the Hilbert space of the current system be $\mathcal{H}_A$, and let us introduce a qubit $\mathcal{H}_B$. Then define 
\begin{equation}
  \begin{aligned}
    \hat{H}_{AB} = \left( \hat{H}_0 + \theta_s \hat{Z} \right) &\otimes I_B + \theta_d \hat{Z} \otimes \sigma_z, \\
    \hat{L}^{AB}_j &= \hat{L}_j \otimes I_B,
  \end{aligned}
\end{equation}
where $\hat{H}_{AB} = \hat{H}_{AB}^{\dag}$ is a legitimate Hermitian Hamiltonian. The master equation describing this is the usual one, 
\begin{equation}\label{eq:master_AB}
  \begin{aligned}
      \partial_t \hat{\rho} \equiv \mathcal{L}^{AB} \hat{\rho} = -i \left[ \hat{H}_{AB}, \hat{\rho} \right] + \sum_{j} \mathcal{D}[\hat{L}_j^{AB}] \hat{\rho}.
  \end{aligned}
\end{equation} 
We formally write the solution as $\hat{\rho}(t) = e^{\mathcal{L}^{AB} t} \hat{\rho}(0)$. Then, it turns out that this conveniently solves Eq.~(\ref{eq:H_2sided_2}) as well, via
\begin{equation}
\hat{\mu}(t) = 2 \langle 0 | e^{\mathcal{L}^{AB} t} \left( \hat{\mu}(0) \otimes | + \rangle \langle + | \right) | 1 \rangle.
\end{equation}  

Let's consider the dephasing of the qubit under Eq.~(\ref{eq:master_AB}). In particular, we are interested in the quantity \begin{equation}
  \begin{aligned}
    e^{-\nu(t)} \equiv 2 {\rm Tr}_A \left[ \left( I_A \otimes \langle 0 | \right) \hat{\rho}(t) \left( I_A \otimes | 1 \rangle \right) \right] = {\rm Tr} \hat{\mu}(t),
  \end{aligned}
\end{equation}
so we see that the trace of the pseudo-density matrix in fact gives exactly the dephasing rate of this system. In this case, the limit $\theta_d \rightarrow 0$ corresponds to the weak coupling limit, described by a noise spectral density. Hence, instead of studying the \textit{metrological properties} of system $A$ directly, one might try to study the \textit{noise spectrum} of system $B$. 

This leads to an intriguing method to calculate the long-time scaling of the Fisher information. We know that the dephasing rate of the qubit in system $A$ is given by the symmetrized noise spectral density at zero frequency (since we don't assume any qubit Hamiltonian), \begin{equation}
  \Gamma = 2 \theta_d^2 S_{ZZ}[0] = 4  \theta_d^2  \frac{1}{2} \int_0^{\infty} dt \langle \{ \hat{Z}(t), \hat{Z}(0) \} \rangle_{A}^{(0)} = 2 \theta_d^2 \int_0^{\infty} dt \langle \{ \hat{Z}(t), \hat{Z}(0) \} \rangle_{A}^{(0)}
\end{equation}
where $\langle \cdot \rangle_A$ is a connected correlator taken with respect to some stationary state of $A$. In absence of any other information, we can take this to be the steady state of system $A$ when $\theta_d = 0$, denoted \begin{equation}
  \langle \cdot \rangle_A^{(0)} \equiv {\rm Tr} \left( (\cdot) \hat{\rho}_{ss, A}^{(0)} \right).
\end{equation} 
For congruence with later calculations, we can now write the integrand as the autocorrelator of the sensor system (A), $C_{ZZ}(t) = \langle \{ \hat{Z}(t), \hat{Z}(0) \} \rangle_{A}^{(0)}$. This allows us to extract the long-time dephasing rates as follows. We identify $\delta = 2 \theta_d$ in Eq.~(\ref{eq:cts_QFI_expression}), and the trace of $\hat{\mu}$ with the dephasing rate in Eq.~(\ref{eq:F_G}), so that we can write 
\begin{equation}\label{app_eq:all_times_I_G}
  \begin{aligned}
    I_G
    &= - 4 \partial_{\delta}^2 \log \left| {\rm Tr} \hat{\mu}_{\theta, \theta + \delta} \right|_{\delta \rightarrow 0} \\
    &= 4 t \partial_{\delta}^2 \Gamma_{\delta \rightarrow 0} \\
    &= t \partial_{\delta}^2 \delta^2 \int d\tau C_{ZZ}(\tau) 
    = 4 \int_0^T d\tau_1 \int_0^{\tau_1} d \tau_2 C_{ZZ}(\tau_2),
  \end{aligned}
\end{equation}
where one may  recall the standard expression for qubit dephasing (see for example Ref.~\cite{Clerk_RMP_2010_noise}). This expression may be obtained from full perturbation theory, as was done for the global QFI in \cite{Gammelmark_PRA_2013_cts_meas_metrology}. 

We would now like to write this in terms of a noise spectral density, which will yield a compact expression for the sensitivity. We may define the symmetrized noise spectrum as 
\begin{equation}
    S_{ZZ}[\omega] \equiv \int_{-\infty}^{\infty} dt C_{ZZ}(t) e^{i \omega t}.
\end{equation}
As $C_{ZZ}(t)$ is necessarily real valued, we have that $S_{ZZ}[\omega] = S_{ZZ}[-\omega]$. Then we may compactly write, in the large $T$ limit, 
\begin{equation}\label{app_eq:zero_freq_I_G}
    I_G \simeq 2 T \int_{- \infty}^{\infty} d \tau C_{ZZ}(\tau) = 2 T S_{ZZ}[0].
\end{equation}
As an aside, simulating the full system with the virtual qubit is much more numerically stable than working directly with the two-sided master equation. This has the cost of halving the system size we can achieve in our simulations, but allows us to compute objects like derivatives and the matrix square root in a much more stable fashion.

\subsection{Filter function approach}

As written, Eq.~(\ref{app_eq:zero_freq_I_G}) is nothing more than a compact way to write the sensitivity in terms of the zero-frequency noise spectral density. Can we use the noise spectrum approach to obtain further physical insight into the problem? One compelling way to re-write the noise spectrum is in terms of filter functions. Given the connection to free dephasing, we can write $I_G(T)$ at any time $T$ in terms of the full noise spectrum $S_{ZZ}[\omega]$ and the standard free-induction decay (Ramsey) filter function $f_G[\omega,T]$:
\begin{equation}\label{app_eq:I_G_filter}
\begin{aligned}
    I_G(T) &= 8 \int_{-\infty}^{\infty} \frac{d \omega}{2 \pi} S_{ZZ}[\omega] f_G[\omega, T], \\
    f_G[\omega, T] &= \frac{\sin^2 ( \omega T/2)}{\omega^2}.
\end{aligned}
\end{equation}  
This standard filter function is peaked at zero frequency, with a width $\propto 1/T$. 
This then tells us that for a fixed $\langle \Delta \hat{Z}^2 \rangle$ the ideal noise spectrum for obtaining a large $I_G$ is one that is sharply peaked at zero.

In the main text, we discussed that for a class of system (i.e.~purely dissipative evolution, steady state either pure or fully mixed), it is also possible to write the environmental QFI $I_E$ solely in terms of the autocorrelation function $C_{ZZ}(t)$.  This expression can also be usefully written in frequency space using an alternate filter function:
\begin{equation}\label{app_eq:I_E_filter}
\begin{aligned}
    I_{\rm E}(T) &= 8 \int_{-\infty}^{\infty} \frac{d \omega}{2 \pi} S_{ZZ}[\omega] f_{\rm E}[\omega, T], \\
    f_{\rm E}[\omega, T] &= \frac{\sin^4 ( \omega T/2)}{\omega^2},
\end{aligned}
\end{equation}
 In contrast to $f_{\rm G}$, the environmental filter function $f_{\rm E}$ is peaked away from zero. We will use this in the next section to derive a bound on $I_E$ for the relevant class of systems.

\subsection{Bounds on the environmental Fisher information}

In this section, we derive some bounds for the environmental Fisher information. We may obtain similar bounds for any $\hat{Z}$ that is a sum of single body terms, but for concreteness and to fix prefactors, we will only consider angular momentum operators, $\hat{Z} = \hat{J}_{\vec{r}}$. For angular momentum operators, we have 
\begin{equation}\label{app_eq:variance_bound}
    \langle \Delta \hat{Z}^2 \rangle \leq \frac{N^2}{4},
\end{equation}
which is saturated by the superposition of minimum and maximum angular momentum states, $| - J_{\vec{r}} \rangle + | J_{\vec{r}} \rangle$.

We will also make use of the fact that for any stationary system, we must have 
\begin{equation}\label{app_eq:C_ZZ_bound}
    C_{ZZ}(t \neq 0) \leq C_{ZZ}(0) = 2 \langle \Delta \hat{Z} \rangle \leq \frac{N^2}{2}.
\end{equation}

We first obtain a loose bound by appealing to the fact that $I_E < I_G$, which follows from the fact that all information in the environment must also be contained in the sum of the environment and the system. Then, using Eqs.~(\ref{app_eq:all_times_I_G}, \ref{app_eq:C_ZZ_bound}), we obtain, 
\begin{equation}
    \begin{aligned}
        I_E \leq I_G = 4 \int_0^T d\tau_1 \int_0^{\tau_1} d \tau_2 C_{ZZ}(\tau_2) \leq 4 C_{ZZ}(0) \int_0^T d\tau_1 \int_0^{\tau_1} d \tau_2 1 = 2 C_{ZZ}(0) T^2 \leq N^2 T^2,
    \end{aligned}
\end{equation}
or \begin{equation}
    I_E \leq N^2 T^2,
\end{equation}
as claimed in the main text.

For the class of systems we consider in the main text, one may obtain a tighter bound using Eq.~(\ref{app_eq:I_E_filter}). The key observation is that $f_{\rm E}$ is peaked at some $\omega_c$, such that 
\begin{equation}
    \begin{aligned}
        f_{\rm E}[\omega, T] \leq &f_{\rm E}[\omega_c, T] \simeq 0.525, \\
        \omega_c &\simeq \frac{2.332}{T},
    \end{aligned}
\end{equation}
which is approximate only insofar as the numerical values are truncated to their third decimal places. Hence, Eq.~(\ref{app_eq:I_E_filter}) is maximized by a some $S_{ZZ}[\omega]$ that is a delta function at $\omega_c$, namely,
\begin{equation}
    S_{ZZ, c}[\omega] = S_0 (\delta(\omega - \omega_c) + \delta(\omega + \omega_c).
\end{equation}
The prefactor $S_0$ may be obtained from the constraint that the integral of $S_{ZZ, c}[\omega]$ should yield $C_{ZZ}(0)$, so that 
\begin{equation}
    S_0 = \pi T \langle \Delta \hat{Z}^2 \rangle \leq \frac{1}{4} \pi T N^2. 
\end{equation}
Hence, 
\begin{equation}
    I_E \leq \frac{2T}{\pi} f_{\rm E}[\omega_c, T] S_0 \leq \frac{1}{2} f_{\rm E}[\omega_c, T] N^2 T^2 \simeq 0.262 N^2 T^2.
\end{equation}

\section{Unnaturally good: Pathological instances of Heisenberg and super-Heisenberg scaling in sensitivity}

Formally, we should think about a system size scaling as a quantity which emerges when we consider a series of systems, say $S_N$, indexed by a system size variable $N$. In this section, we show that one may construct systematically series of systems where the sensitivity grows as $O(N^2)$, whereas the finite-time environmental QFI does not. This demonstrates that the Heisenberg scaling of the sensitivity is in general not a good indicator of a Heisenberg scaling in the QFI.

For our first example, we consider a simple qubit system, where $S_1$ comprises a qubit subject to loss, 
\begin{equation}
    S_1: \partial_t \hat{\rho} = \Gamma_1 \mathcal{D}[\hat{\sigma}_-] \hat{\rho},
\end{equation}
where the subscript of $\Gamma_1$ follows the subscript of $S_1$ and $\Gamma_1 = \Gamma$ is a constant. It is straightforward to show, via exact diagonalization or otherwise, that the environmental QFI for some fixed integration time $T$ goes like 
\begin{equation}
 (I_E^{\Gamma})^{(1)}[\hat{\sigma}_x](T) = \frac{4}{\Gamma} T + \frac{16}{\Gamma^2} e^{- \Gamma T / 2} - \frac{4}{\Gamma^2} e^{- \Gamma T} - \frac{12}{\Gamma^2},
\end{equation}
where we can read off the sensitivity as $\mathcal{S}_1 = 4/\Gamma$. We now consider a series of systems given by 
\begin{equation}
    S_N: \partial_t \hat{\rho} = \Gamma_N \sum_{i=1}^N \mathcal{D}[ \hat{\sigma}_-^{(i)}] \hat{\rho},
\end{equation}
where we set $\Gamma_N = \Gamma / N$. Since this involves the uncorrelated evolution of $N$ qubits, we have 
\begin{equation}
\begin{aligned}
    (I_E^{\Gamma})^{(N)}\left[\sum_{i=1}^N \hat{\sigma}_x^{(i)}\right](T) 
    &= \sum_{i=1}^N (I_E^{\Gamma})^{(N)}\left[ \hat{\sigma}_x^{(i)}\right](T) \\
    &= N (I_E^{\Gamma})^{(N)}\left[ \hat{\sigma}_x^{(i)}\right](T) \\
    &= 
    N \left( \frac{4}{\Gamma_N} T + \frac{16}{\Gamma^2_N} e^{- \Gamma_N T / 2} - \frac{4}{\Gamma^2_N} e^{- \Gamma_N T} - \frac{12}{\Gamma^2_N} \right), \\
    &= \frac{4N^2}{\Gamma} T + \frac{16N^3}{\Gamma^2} e^{- \Gamma T / 2 N} - \frac{4N^3}{\Gamma^2} e^{- \Gamma T/N} - \frac{12 N^4}{\Gamma^2}, 
\end{aligned}
\end{equation}
with the sensitivity displaying `Heisenberg scaling', since for this series of systems, $\mathcal{S}_N = 4 N^2 / \Gamma \sim O(N)$. However, when $N$ becomes large, we have 
\begin{equation}
    (I_E^{\Gamma})^{(N)}\left[\sum_{i=1}^N \hat{\sigma}_x^{(i)}\right](T) \rightarrow \frac{1}{3N} \Gamma T^3 \rightarrow 0,
\end{equation}
where the limit holds when we keep $T$ fixed and increase $N$. Hence, despite a sensitivity that displays Heisenberg scaling, the actual finite-time environmental QFI goes to $0$.

While the series of systems $S_N$ in this example appears contrived, we note that this also shows up in other more natural settings. In particular, one may consider the boundary time crystal of \cite{Cabot_PRL_2024_BTC_sensing}, but using the usual parameterization with a Kac factor,
\begin{equation}
    S_N: \partial_t \hat{\rho} = -i \Omega [\hat{J_x}, \hat{\rho}] + \frac{1}{N} \Gamma \mathcal{D}[\hat{J}_-] \hat{\rho},
\end{equation}
with $\Omega > 2 \Gamma$. Then, since the parameterization without the Kac factor yields a sensitivity that goes like $N^2/\Gamma$, one finds that for this parameterization, we will obtain $\mathcal{S}_N \sim O(N^3)$, which appears like a `super-Heisenberg' scaling -- this is obtained for example in Ref.~\cite{Midha_arxiv_2025_metrology_open}. However, similarly the actual environmental QFI goes to $0$ for large $N$ and fixed $T$.


\section{Environmental QFI for high-temperature superradiant sensor}

\subsection{Setup}

The high-temperature sensor has a Lindbladian given by 
\begin{equation}
\begin{aligned}
    \partial_t \hat{\rho} = \mathcal{L} \hat{\rho} = \Gamma \sum_{\alpha = \pm} \mathcal{D}[\hat{J}_{\alpha}] \hat{\rho},
\end{aligned}
\end{equation}
with the property that $\mathcal{L} = \mathcal{L}^{\dag}$ with respect to the Hilbert-Schmidt inner product. In situations where ambiguity may arise, we may put the operator a given superoperator acts on in square brackets eg. $\mathcal{L} \hat{\rho} = \mathcal{L} [\hat{\rho}]$. For notational convenience, we define the super operators
\begin{equation}
    \begin{aligned}
        \mathcal{Z}^{R} [ \cdot ]= i ( \cdot ) \hat{Z}, 
        \mathcal{Z}^{L} [ \cdot ]= - i \hat{Z} ( \cdot ).
    \end{aligned}
\end{equation}
Then, the two-sided master equation may be written,
\begin{equation}\label{app_eq:superoperator_TSME}
    \begin{aligned}
        \partial_t \hat{\mu} = (\mathcal{L} + \theta \mathcal{Z}^R)[\hat{\mu}], \\
        \partial_t \hat{\mu}^{\dag} = (\mathcal{L} + \theta \mathcal{Z}^L)[\hat{\mu}^{\dag}].
    \end{aligned}
\end{equation}

\subsection{QFI for general operators}

We claim that the environmental QFI for an arbitrary generator $\hat{Z}$ in the high-temperature system is given by, 
\begin{equation}\label{app_eq:inf_temp_I_E}
    \begin{aligned}
        I_{\rm E}(T) = 4 \int_0^T d \tau_1 \int_0^{\tau_1} d \tau_2 C_{ZZ}(\tau_2)
        - 4 \int_0^T d \tau_1 \int_0^{\tau_1} d \tau_2 C_{ZZ}(\tau_1 + \tau_2),
    \end{aligned}
\end{equation}
when $\hat{\mu}(0) = \hat{\rho}_0 = \hat{I}/D$ is the steady state, with $D$ being the dimension of the system. The maximally mixed steady state allows us to greatly simplify the calculation of $I_{\rm E}(T)$. In this case, since the matrix square root is perturbative in $\theta$, we may take it simply as say $\sqrt{\hat{I} + \theta^2 \hat{M}} = \hat{I} + (1/2) \theta^2 \hat{M} + O(\theta^3)$ for a Hermitian operator $\hat{M}$. Since we only require up to $O(\theta^2)$ for the calculation of $I_{\rm E}$, this will yield exact analytical results for the QFI. For a general steady state, one must resort to methods such as the Daleckii-Krein theorem \cite{Daletskii_Krein_AMS_1965, Carlsson_arxiv_2018_daleckii_ext} to perturbatively take the matrix square root.

Our derivation of Eq.~(\ref{app_eq:inf_temp_I_E}) will depend crucially on two assumptions:
\begin{enumerate}
    \item The steady state is the maximally mixed state.
    \item The Lindbladian is self-adjoint.
\end{enumerate}
Beyond the high-temperature model, highly mixed steady states are highly common; at any rate for any continuous sensing protocol, one may put in any steady state, including the maximally mixed state. Using just assumption (1), we first arrive at the intermediate result Eq.~(\ref{app_eq:intermediate_I_E}). We must then use assumption (2) to arrive at the final Eq.~(\ref{app_eq:inf_temp_I_E}). 


We will begin by formally perturbing the the propagator associated with the two-sided master equation, $\mathcal{V}^R(t)$, satisfying $\hat{\mu}(T) = \mathcal{V}^R(T) \hat{\mu}(0)$, as a Dyson series \cite{Sakurai_Napolitano_2021}. Keeping terms up to second order in $\theta$,
\begin{equation}
    \mathcal{V}^R(T) = \mathcal{I} + \theta \int_0^T dt_1 e^{\mathcal{L}(T - t_1) } \circ \mathcal{Z}^R \circ e^{\mathcal{L} t_1 }  + \theta^2 \int_0^T dt_1 \int_0^{t_1} dt_2 e^{\mathcal{L}(T - t_1) } \circ \mathcal{Z}^R \circ e^{\mathcal{L} (t_1 - t_2) } \circ \mathcal{Z}^R \circ e^{\mathcal{L} t_2 } + O(\theta^3),
\end{equation}
where $\circ$ indicates the composition of superoperators, and $\mathcal{I}$ is the identity superoperator.

Taking the initial state to be the steady state, $\hat{\mu}(0) = \hat{\rho}_0 = e^{\mathcal{L} t } \hat{\rho}_0$, and using $\hat{\mu}(T) = \mathcal{V}^R(T) \hat{\mu}(0)$, we obtain, 
\begin{equation}\label{app_eq:full_ptb_0}
    \hat{\mu}(T) = \hat{\rho}_0 + \theta \int_0^T dt_1 e^{\mathcal{L}t_1} \circ \mathcal{Z}^R [\hat{\rho}_0]  + \theta^2 \int_0^T dt_1 \int_0^{t_1} dt_2 e^{\mathcal{L} t_1 } \circ \mathcal{Z}^R \circ e^{\mathcal{L} (t_1 - t_2) } \circ \mathcal{Z}^R [\hat{\rho}_0] + O(\theta^3),
\end{equation}
where we have redefined the limits of integration to remove $T$'s from the exponents. We proceed similarly to obtain a similar expression for $\hat{\mu}^{\dag}(T)$, with the only difference being $\mathcal{Z}^R \rightarrow \mathcal{Z}^L$. Multiplying them, 
\begin{equation}\label{app_eq:full_ptb_1}
    \begin{aligned}
        \hat{\mu}(t) \hat{\mu}(t)^{\dag} 
        = \hat{\rho}_0^2 &+ 
        \theta \int_0^T dt_1 \left( \hat{\rho}_0 e^{\mathcal{L} t_1 } \circ  \mathcal{Z}^R [\hat{\rho}_0] + e^{\mathcal{L} t_1 } \circ  \mathcal{Z}^L [\hat{\rho}_0] \hat{\rho}_0 \right) \\
        &+ \theta^2 \int_0^T d t_1 \int_0^T dt_2   e^{\mathcal{L}t_1} \circ \mathcal{Z}^R [\hat{\rho}_0]  e^{\mathcal{L}t_2} \circ \mathcal{Z}^L [\hat{\rho}_0] \ \\
        &+ \theta^2 \int_0^T d t_1 \int_0^{t_1} dt_2 \left( e^{\mathcal{L} t_1 } \circ \mathcal{Z}^R \circ e^{\mathcal{L} (t_1 - t_2) } \circ \mathcal{Z}^R [\hat{\rho}_0] \hat{\rho}_0 + \hat{\rho}_0 e^{\mathcal{L} t_1 } \circ \mathcal{Z}^L \circ e^{\mathcal{L} (t_1 - t_2) } \circ \mathcal{Z}^L [\hat{\rho}_0] \right).
    \end{aligned}
\end{equation}

We now specialize to the high-temperature model and set $\hat{\rho}_0 = I/D$. The $O(\theta)$ term disappears, since 
\begin{equation}
    \begin{aligned}
    \int_0^T dt_1 \left( \hat{\rho}_0 e^{\mathcal{L} t_1 } \circ  \mathcal{Z}^R [\hat{\rho}_0] + e^{\mathcal{L} t_1 } \circ  \mathcal{Z}^L [\hat{\rho}_0] \hat{\rho}_0 \right) 
    &= \frac{1}{D^2 }\int_0^T dt_1  e^{\mathcal{L} t_1 } \circ  (\mathcal{Z}^R + \mathcal{Z}^L) [I] \\
    &= \frac{1}{D^2 }\int_0^T dt_1  e^{\mathcal{L} t_1 } [i \hat{Z} - i \hat{Z}] = 0.
\end{aligned}
\end{equation}
Since the $O(\theta^2)$ term is Hermitian, and $\hat{\rho}_0^2 = \hat{I}/D^2$, we may expand $\mathcal{Z}^{L/R}$ and take the square root perturbatively to obtain
\begin{equation}\label{app_eq:full_ptb_2}
    \begin{aligned}
        &\sqrt{\hat{\mu}(T) \hat{\mu}(T)^{\dag} }
        = \mathcal{I} \\
        &\qquad + \frac{1}{2} D \theta^2 {\rm Tr} \int_0^T d t_1 \int_0^T dt_2   e^{\mathcal{L}t_1} \circ \mathcal{Z}^R [\hat{\rho}_0]  e^{\mathcal{L}t_2} \circ \mathcal{Z}^L [\hat{\rho}_0] \ \\
        &\qquad + \frac{1}{2} \theta^2 D {\rm Tr}  \int_0^T d t_1 \int_0^{t_1} dt_2 \left( e^{\mathcal{L} t_1 } \circ \mathcal{Z}^R \circ e^{\mathcal{L} (t_1 - t_2) } \circ \mathcal{Z}^R [\hat{\rho}_0] \hat{\rho}_0 + \hat{\rho}_0 e^{\mathcal{L} t_1 } \circ \mathcal{Z}^L \circ e^{\mathcal{L} (t_1 - t_2) } \circ \mathcal{Z}^L [\hat{\rho}_0] \right).
    \end{aligned}
\end{equation}

The first term in the integrand of the second line in Eq.~(\ref{app_eq:full_ptb_2}) maybe written
\begin{equation}\label{app_eq:simplified_second_line_1}
    \begin{aligned}
        {\rm Tr} \left( e^{\mathcal{L}t_1} \circ \mathcal{Z}^R [\hat{\rho}_0]  e^{\mathcal{L}t_2} \circ \mathcal{Z}^L [\hat{\rho}_0] \right) 
        &= \frac{1}{D} {\rm Tr} \left( e^{\mathcal{L}t_1}[ \hat{\rho}_0 \hat{Z}]  e^{\mathcal{L}t_2} [ \hat{Z} ] \right) \\
        &= \frac{1}{D} {\rm Tr} \left( \hat{\rho}_0 \hat{Z} e^{ \mathcal{L}^{\dag} t_2 } \circ e^{\mathcal{L} t_1 } [\hat{Z}] \right) = \frac{1}{D} \left\langle \hat{Z} e^{ \mathcal{L}^{\dag} t_2 } \circ e^{\mathcal{L} t_1 } \hat{Z} \right\rangle^{(0)},
    \end{aligned}
\end{equation}
where we have used the definition of the adjoint of the Lindbladian superoperator under the Hilbert-Schmidt inner product, and in the final equality, we denote by $\langle ( \cdot) \rangle^{(0)}$ an expectation value taken with respect to the steady state. The integrand of the third line in Eq.~(\ref{app_eq:full_ptb_2}) may be similarly simplified; focusing on the term containing $\mathcal{Z}^R$, 
\begin{equation}\label{app_eq:simplified_third_line_1}
    \begin{aligned}
        {\rm Tr}  \left( e^{\mathcal{L} t_1 } \circ \mathcal{Z}^R \circ e^{\mathcal{L} (t_1 - t_2) } \circ \mathcal{Z}^R [\hat{\rho}_0] \hat{\rho}_0 \right) 
        &= - \frac{1}{D}{\rm Tr} \left(  e^{\mathcal{L} (t_1 - t_2) } [ \hat{Z}] \hat{Z}  e^{\mathcal{L}^{\dag} t_1} {\hat{\rho}_0} \right) \\
        &= - \frac{1}{D}{\rm Tr} \left( e^{\mathcal{L} (t_1 - t_2) } [  \hat{Z}] \hat{Z} \hat{\rho}_0 \right) \\
        &= - \frac{1}{D} {\rm Tr} \left( e^{\mathcal{L} (t_1 - t_2) } [ \hat{Z}] \hat{Z} \hat{\rho}_0 \right) 
        = - \frac{1}{D} \langle \hat{Z} \hat{Z}(t_1 - t_2) \rangle^{(0)},
    \end{aligned}
\end{equation}
where we note that the second equality follows from the fact that $e^{\mathcal{L}^{\dag} t_1 }[\hat{I}] = \hat{I}$, which in turn follows from the fact that $\mathcal{L}$ is trace-preserving. The full integrand of the second line of Eq.~(\ref{app_eq:full_ptb_2}) is obtained by symmetrizing this quantity, 
\begin{equation}\label{app_eq:symmetrized_third_line_1}
    \begin{aligned}
        - \frac{1}{D} \langle \hat{Z} \hat{Z}(t_1 - t_2) \rangle^{(0)} - \frac{1}{D} \langle \hat{Z} (t_1 - t_2) \hat{Z} \rangle^{(0)} = - \frac{1}{D} C_{ZZ}(t_1 - t_2),
    \end{aligned}
\end{equation}
where we can now recognize the emergence of the autocorrelator.

At this stage, we have only used assumption (1); we may plug Eq.~(\ref{app_eq:simplified_second_line_1}, \ref{app_eq:symmetrized_third_line_1}) into Eq.~(\ref{app_eq:full_ptb_2}) to obtain, 
\begin{equation}\label{app_eq:full_ptb_3}
    \begin{aligned}
        \sqrt{\hat{\mu}(T) \hat{\mu}(T)^{\dag} }
        = 1 
        - \frac{1}{2} \theta^2 \int_0^T d t_1 \int_0^{t_1} dt_2 C_{ZZ}(t_2) + 
        + \theta^2 \int_0^T d t_1 \int_0^{t_2} dt_2  \left\langle \hat{Z} e^{ \mathcal{L}^{\dag} t_2 } \circ e^{\mathcal{L} t_1 } \hat{Z} \right\rangle^{(0)} + O(\theta^3),
    \end{aligned}
\end{equation} 
where we have re-defined integration variables in the first integrand to keep only one term in the argument of $C_{ZZ}$. We may differentiate to obtain the following intermediate expression for the environmental QFI, 
\begin{equation}\label{app_eq:intermediate_I_E}
    \begin{aligned}
        I_{\rm E}(T) = 4 \int_0^T d \tau_1 \int_0^{\tau_1} d \tau_2 C_{ZZ}(\tau_2) - 4 \int_0^T d \tau_1 \int_0^{T} d \tau_2  \left\langle \hat{Z} e^{ \mathcal{L}^{\dag} \tau_2 } \circ e^{\mathcal{L} \tau_1 } \hat{Z} \right\rangle^{(0)},
    \end{aligned}
\end{equation}
which is valid for any system where the steady state is (or may be approximated by) the maximally mixed state.

Finally, we can obtain the result for the high-temperature model by noting that since $\mathcal{L} = \mathcal{L}^{\dag}$, we have $ \left\langle \hat{Z} e^{ \mathcal{L}^{\dag} t_2 } \circ e^{\mathcal{L} t_1 } \hat{Z} \right\rangle^{(0)} =  \left\langle \hat{Z} e^{\mathcal{L} (t_1 + t_2) } \hat{Z} \right\rangle^{(0)}$. Plugging this into Eq.~(\ref{app_eq:intermediate_I_E}) and symmetrizing to obtain the autocorrelator $C_{ZZ}$, the final result Eq.~(\ref{app_eq:inf_temp_I_E}) is obtained. With that we can now solve for $C_{ZZ}$ in order to obtain explicit expressions for Eq.~(\ref{app_eq:inf_temp_I_E}).

\subsection{Autocorrelators for spin operators}

We start with $\hat{J}_z$. The adjoint equation gives 
\begin{equation}
    \mathcal{L}^{\dag}(\hat{J}_z) = \Gamma (\hat{J}_- \hat{J}_+ - \hat{J}_+ \hat{J}_-) = - 2 \Gamma \hat{J}_z,
\end{equation}
from which it follows, using the quantum regression theorem and making a stationary approximation,
\begin{equation}
    \begin{aligned}
        \partial_t C_{J_z J_z}(t) 
        &= \langle \mathcal{L}^{\dag}(\hat{J}_z) \hat{J}_z \rangle + \langle \hat{J}_z \mathcal{L}^{\dag}(\hat{J}_z) \rangle - 2 \langle \mathcal{L}^{\dag}(\hat{J}_z) \rangle \langle \hat{J}_z \rangle \\
        &= - 2 \Gamma C_{J_z J_z}(t) \implies C_{J_z J_z}(T) = 2 \langle \Delta \hat{J}_z^2 \rangle e^{- 2 \Gamma T}.
    \end{aligned}
\end{equation}

We perform a similar computation for $\hat{J}_x$, 
\begin{equation}
    \begin{aligned}
        \frac{1}{\Gamma} \mathcal{L}^{\dag}(\hat{J}_x) 
        &= \frac{1}{2} [\hat{J}_-, \hat{J}_x] \hat{J}_+ + \frac{1}{2} \hat{J}_- [\hat{J}_x, \hat{J}_+] + \frac{1}{2}[\hat{J}_+, \hat{J}_x] \hat{J}_- + \frac{1}{2} \hat{J}_+ [\hat{J}_x, \hat{J}_-] \\
        &= \frac{1}{2} [\hat{J}_+, \hat{J}_z] + \frac{1}{2} [\hat{J}_z, \hat{J}_-] \\
        &= - \frac{1}{2}\hat{J}_+ - \frac{1}{2} \hat{J}_- = - \hat{J}_x,
    \end{aligned}
\end{equation}
from which it follows that 
\begin{equation}
    \begin{aligned}
        \partial_t C_{J_x J_x}(t) = - \Gamma C_{J_x J_x}(t) \implies C_{J_x J_x}(T) = 2 \langle \Delta \hat{J}_x^2 \rangle e^{- \Gamma T},
    \end{aligned}
\end{equation}
with a similar result holding for $\hat{J}_y$. The optimal values of the environmental QFI follow.

\subsection{Optimal measurement for the high-temperature sensor}

In this section, we explicitly construct a decoder for the high-temperature sensor. Note that since this is a thermal system, a priori we know that a simple absorber exists. 

In particular, let $\hat{O}^{(A)}$ denote an operator on our original sensor system, and $\hat{O}^{(B)}$ denote an operator on the downstream system. Then, the following cascaded system \cite{Gardiner_PRL_1993_csc_1, Carmichael_PRL_1993_csc_2}, specified by Hamiltonian and jump operators, 
\begin{equation}
    \begin{aligned}
        \hat{H}_{\rm csc} &= - \frac{i}{2} \left( \hat{J}^{(A)}_- \otimes \hat{J}^{(B)}_+ - \hat{J}^{(A)}_+ \otimes \hat{J}^{(B)}_- \right)\\
        \hat{L}_1 &= \hat{J}^{(A)}_+ \otimes \hat{I}^{(B)} - \hat{I}^{(A)} \otimes \hat{J}^{(B)}_-, \\
        \hat{L}_2 &= \hat{J}^{(A)}_- \otimes \hat{I}^{(B)} - \hat{I}^{(A)} \otimes \hat{J}^{(B)}_+, 
    \end{aligned}
\end{equation}
hosts the dark state 
\begin{equation}
    | \Psi \rangle = \frac{1}{\sqrt{2J+1}} \sum_{n = - J}^{n = J} |J, n \rangle^{(A)} \otimes |J, n \rangle^{(B)},
\end{equation}
where $|J, n \rangle$ enumerates the eigenstates of $\hat{J}_z$. It is straightforward to check that $\hat{H}_{\rm csc} | \Psi \rangle = \hat{L}_{1, 2} | \Psi \rangle = 0$, and ${\rm Tr}_B | \Psi \rangle \langle \Psi | = \hat{I}/2J+1$. We note that this solution is not unique, and may be modified with additional phases on the system $B$ operators.

\subsection{Beyond the high-temperature sensor}

We may generalize the above result by considering the following set of ``completely dephasing'' models, where 
\begin{equation}
    \partial_t \hat{\rho} = \mathcal{L} \hat{\rho} = 2 \Gamma (1 + \eta) \mathcal{D}[\hat{J}_x] \hat{\rho} + 2 \Gamma (1 - \eta) \mathcal{D}[\hat{J}_y] \hat{\rho},
\end{equation}
which satisfy both the conditions (1) and (2). When $\eta = 0$, this is exactly the high-temperature model; we can think about $\eta$ as a parameter controlling the bias of the dephasing. The autocorrelator EOM of this model is given by 
\begin{equation}
    \begin{aligned}
        C_{J_z J_z}(T) &= 2 \langle \Delta \hat{J}_z^2 \rangle^{(0)} e^{- 2 \Gamma T}, \\ 
        C_{J_x J_x}(T) &= 2 \langle \Delta \hat{J}_x^2 \rangle^{(0)} e^{- \Gamma (1 - \eta) T}, \\
        C_{J_y J_y}(T) &= 2 \langle \Delta \hat{J}_y^2 \rangle^{(0)} e^{- \Gamma (1 + \eta) T}.
    \end{aligned}
\end{equation}
Since the autocorrelator takes the form of a simple exponential decay, all conclusions from the main text regarding $I_E^{\rm opt}$ of the high-temperature model hold in an almost identical way. These class of models also show that the Heisenberg scaling of the high-temperature model is robust against imbalances in the $x, y$ dephasing rates.

\section{Environmental QFI for Dissipative Spin-Squeezer}

\subsection{Setup}

The dissipative spin-squeezer has a Lindbladian given by,
\begin{equation}
    \partial_t \hat{\rho} = \Gamma \mathcal{D}[\hat{J}_+ - \tanh(r) \hat{J}_-] \hat{\rho},
\end{equation}
where $r$ is a squeezing parameter -- the larger $r$ is, the more spin-squeezed the steady state, which is a pure dark state of the system for even $N$ \cite{Groszkowski_PRX_2022_dissipative_spin_squeezing}.

\subsection{QFI for general operators}

We claim that the environmental QFI for an arbitrary generator $\hat{Z}$ in the dissipative spin-squeezer is given by, 
\begin{equation}\label{app_eq:spin_sq_I_E}
    \begin{aligned}
        I_E(T) = 4 \int_0^T d \tau_1 \int_0^{\tau_1} d \tau_2 C_{ZZ}(\tau_2)
        - 4 \int_0^T d \tau_1 \int_0^{\tau_1} d \tau_2 C_{ZZ}(\tau_1 + \tau_2),
    \end{aligned}
\end{equation}
which is identical to Eq.~(\ref{app_eq:inf_temp_I_E}). In contrast to the high-temperature model, the key insight here is that for a system with a pure dark state, $\hat{\mu}(T) \hat{\mu}(T)^{\dag}$ will be rank $1$, hence, we may take its square root as one would do a scalar. Similar to the derivation of Eq.~(\ref{app_eq:inf_temp_I_E}), the derivation here depends on two facts about the dissipative spin squeezer:
\begin{enumerate}
    \item The steady state is a pure dark state.
    \item The system is dissipation only.
\end{enumerate}
The first fact is interesting for two reasons: First, there are large classes of Lindbladians with pure dark states, and second, for any such sensor, the optimal measurement is simply photodetection on the output field (see Sec.~\ref{app_sec:photodetection}). Hence, we will proceed by just assuming (1) at first, deferring (2) to the last possible moment.

We begin by considering a system with both a Hamiltonian $\hat{H}_0$ and dissipative terms $\{\hat{L}_i\}$. Assuming that the system hosts a dark state $| \Psi \rangle$, we may write the steady state as $\hat{\rho}_{0} = | \Psi \rangle \langle \Psi |$, where $|\Psi \rangle$ satisfies (up to a possible shift of $\hat{H}_0$), 
\begin{equation}\label{app_eq:dark_state_cond}
    \begin{aligned}
        \hat{H}_0 | \Psi \rangle &= 0, \\
        L_i | \Psi \rangle &= 0, \quad \forall \quad i.
    \end{aligned}
\end{equation} 
We may then also define the effective Hamiltonian, 
\begin{equation}\label{app_eq:H_eff}
    \hat{H}_{\rm eff} = \hat{H}_0 - \frac{1}{2} i \sum_j L^{\dag}_j L_j.
\end{equation}

We may start from Eq.~(\ref{app_eq:full_ptb_0}). In the case when $\hat{\rho}_0 = | \Psi \rangle \langle \Psi |$ is a dark state, this may be written simply in terms of the effective Hamiltonian as 
\begin{equation}\label{app_eq:dyson_H_eff}
    \hat{\mu}(T) = | \Psi \rangle \langle \Psi | \left( 1 + i \theta \int_0^T dt_1 \hat{Z} e^{i \hat{H}_{\rm eff}^{\dag} t_1} - \theta^2 \int_0^T dt_1 \int_0^{t_1} dt_2 \hat{Z} e^{i \hat{H}_{\rm eff}^{\dag} (t_1 - t_2)} \hat{Z} e^{i \hat{H}_{\rm eff}^{\dag} t_1} \right) + O(\theta^3),
\end{equation}
where we have simply used the fact that $\hat{H}_{\rm eff}$ or $\hat{L}_i$ acting from the left has trivial action. Next, Eq.~(\ref{app_eq:full_ptb_1}) becomes 
\begin{equation}\label{app_eq:full_ptb_1b}
    \begin{aligned}
        \hat{\mu}(T) \hat{\mu}(T)^{\dag} 
        = &| \Psi \rangle \langle \Psi | \left( 1 + i \theta \int_0^T dt_1 \hat{Z} e^{i \hat{H}_{\rm eff}^{\dag} t_1} - \theta^2 \int_0^T dt_1 \int_0^{t_1} dt_2 \hat{Z} e^{i \hat{H}_{\rm eff}^{\dag} (t_1 - t_2)} \hat{Z} e^{i \hat{H}_{\rm eff}^{\dag} t_1} \right) \\
        &\times \left( 1 - i \theta \int_0^T dt_1 e^{- i \hat{H}_{\rm eff} t_1} \hat{Z}  - \theta^2 \int_0^T dt_1 \int_0^{t_1} dt_2 e^{- i \hat{H}_{\rm eff} t_1} \hat{Z} e^{-i \hat{H}_{\rm eff} (t_1 - t_2)} \hat{Z}  \right) | \Psi \rangle \langle \Psi | \\
        &= | \Psi \rangle \langle \Psi | 
        + \theta^2 | \Psi \rangle \langle \Psi |  \int_0^T d t_1 \int_0^T dt_2  \hat{Z} e^{i \hat{H}_{\rm eff}^{\dag} t_1} e^{- i \hat{H}_{\rm eff} t_2} \hat{Z} | \Psi \rangle \langle \Psi |   \\
        &+ \theta^2 | \Psi \rangle \langle \Psi |   \int_0^T d t_1 \int_0^{t_1} dt_2 \left( \hat{Z} e^{i \hat{H}_{\rm eff}^{\dag} (t_1 - t_2)} \hat{Z}  + \hat{Z} e^{-i \hat{H}_{\rm eff} (t_1 - t_2)} \hat{Z} \right) | \Psi \rangle \langle \Psi |  \\
        &= | \Psi \rangle \langle \Psi | 
        + \theta^2 | \Psi \rangle \langle \Psi |  \int_0^T d t_1 \int_0^T dt_2 \left\langle \hat{Z} e^{i \hat{H}_{\rm eff}^{\dag} t_1} e^{- i \hat{H}_{\rm eff} t_2} \hat{Z} \right\rangle   \\
        &- \theta^2 | \Psi \rangle \langle \Psi |   \int_0^T d t_1 \int_0^{t_1} dt_2 \left( \left\langle \hat{Z} e^{i \hat{H}_{\rm eff}^{\dag} (t_1 - t_2)} \hat{Z} \right\rangle + \left\langle \hat{Z} e^{-i \hat{H}_{\rm eff} (t_1 - t_2)} \hat{Z} \right\rangle \right).
    \end{aligned}
\end{equation}
Since this is a rank $1$ object, the square root is simply taken like a scalar, 
\begin{equation}\label{app_eq:diss_ss_sqrt}
    \begin{aligned}
        {\rm Tr} \sqrt{\hat{\mu}(t) \hat{\mu}(t)^{\dag}}
        &= 1 
        + \frac{1}{2} \theta^2 \int_0^T d t_1 \int_0^T dt_2 \left\langle \hat{Z} e^{i \hat{H}_{\rm eff}^{\dag} t_1} e^{- i \hat{H}_{\rm eff} t_2} \hat{Z} \right\rangle   \\
        &- \frac{1}{2} \theta^2 \int_0^T d t_1 \int_0^{t_1} dt_2 \left( \left\langle \hat{Z} e^{i \hat{H}_{\rm eff}^{\dag} (t_1 - t_2)} \hat{Z} \right\rangle + \left\langle \hat{Z} e^{-i \hat{H}_{\rm eff} (t_1 - t_2)} \hat{Z} \right\rangle \right).
    \end{aligned}
\end{equation}
Recognizing the auto-correlator in the second line, we may take the derivative with respect to $\theta$ to yield the intermediate expression,
\begin{equation}\label{app_eq:intermediate_I_E_ss}
    \begin{aligned}
        I_{\rm E}(T) = 4 \int_0^T d \tau_1 \int_0^{\tau_1} d \tau_2 C_{ZZ}(\tau_2) - 4 \int_0^T d \tau_1 \int_0^{T} d \tau_2  \left\langle \hat{Z} e^{i \hat{H}_{\rm eff}^{\dag} \tau_1} e^{- i \hat{H}_{\rm eff} \tau_2}  \hat{Z} \right\rangle^{(0)},
    \end{aligned}
\end{equation}
which is valid for \textit{any system hosting a dark state}, and despite relying on a very different assumption is in fact identical to Eq.~(\ref{app_eq:intermediate_I_E}).

Finally, we can use fact (2) to specialize to the dissipative spin-squeezing problem for which the $\hat{H}_{\rm eff} = -\hat{H}_{\rm eff}^{\dag}$. We note that the dissipative spin-squeezer aside, this is also valid for any dissipation-only system. Using this in Eq.~(\ref{app_eq:intermediate_I_E_ss}) yields the final expression Eq.~(\ref{app_eq:spin_sq_I_E}). Explicitly, the last term in Eq.~(\ref{app_eq:intermediate_I_E_ss}) reduces as,
\begin{equation}
    \begin{aligned}
        &\int_0^T d \tau_1 \int_0^{T} d \tau_2  \left\langle \hat{Z} e^{i \hat{H}_{\rm eff}^{\dag} \tau_1} e^{- i \hat{H}_{\rm eff} \tau_2}  \hat{Z} \right\rangle^{(0)}\\
        &= 
    \end{aligned}
\end{equation}

\subsection{Autocorrelators for spin operators}

We now compute an approximation for $C_{J_x J_x}$ using a cumulant expansion. It will be useful to note that \begin{equation}
  \hat{L} = (1 - \tanh(r)) \hat{J}_x - i (1 + \tanh(r)) \hat{J}_y.
\end{equation}

First, the relevant commutators that turn up are \begin{equation}
  \begin{aligned}
    \left[ \hat{J}_x, \hat{L} \right] 
    &= \left[ \hat{J}_x, \hat{J}_- \right] - \tanh(r) \left[ \hat{J}_x, \hat{J}_+ \right] 
    = (1 + \tanh(r)) \hat{J}_z = \left( \left[ \hat{L}^{\dag} \hat{J}_x \right] \right)^{\dag}, \\
    \left[ \hat{J}_y, \hat{L}  \right] 
    &= \left[ \hat{J}_y, \hat{J}_- \right] - \tanh(r) \left[ \hat{J}_y, \hat{J}_+ \right] 
    = - i (1 - \tanh(r)) \hat{J}_z = \left( \left[ \hat{L}^{\dag}, \hat{J}_y \right] \right)^{\dag},
  \end{aligned}
\end{equation}

Now, using the quantum regression theorem, we calculate the EOM for $\hat{J}_{x}$. 
\begin{equation}
  \begin{aligned}
    \frac{1}{\Gamma} \partial_t \hat{J}_x 
    &= \frac{1}{2} (1 - \tanh(r)^2) (\hat{J}_z \hat{J}_x + \hat{J}_x \hat{J}_z)  
    - \frac{1}{2} (1 + \tanh(r))^2  \hat{J}_x.
  \end{aligned}
\end{equation}
Now, we can plug this into the two-time correlator to get EOMs.
\begin{equation}\label{app_eq:diss_ss_eom_1}
  \begin{aligned}
    &\frac{1}{\Gamma }\partial_t C_{J_x J_x}(t) 
    = \langle (\partial_t \hat{J}_x(t)) \hat{J}_x(0) \rangle - \langle \partial_t \hat{J}_x(t) \rangle \langle \hat{J}_x(0) \rangle \\
    &= \frac{1}{2} (1 - \tanh(r)^2) (\langle \hat{J}_z(t) \hat{J}_x(t) \hat{J}_x(0) \rangle + \langle \hat{J}_x(t) \hat{J}_z(t) \hat{J}_x(0) \rangle)  
    - \frac{1}{2} (1 + \tanh(r))^2 C_{J_x J_x}(t) \\
    &\qquad - \frac{1}{2} (1 - \tanh(r)^2) (\langle \hat{J}_z(t) \hat{J}_x(t) \rangle \langle \hat{J}_x(0) \rangle + \langle \hat{J}_x(t) \hat{J}_z(t) \rangle \langle \hat{J}_x(0) \rangle)  
  \end{aligned}
\end{equation}

Applying a second-order cumulant expansion \cite{Kubo_1962}, we have 
\begin{equation}
  \begin{aligned}
    \langle \hat{J}_z(t) \hat{J}_x(t) \hat{J}_x(0) \rangle - \langle \hat{J}_z(t) \hat{J}_x(t) \rangle \langle \hat{J}_x(0) \rangle
    &\simeq \langle \hat{J}_z(t) \rangle C_{J_x J_x}(t) + \langle \hat{J}_x(t) \rangle C_{J_z J_x}(t), \\
    \langle \hat{J}_x(t) \hat{J}_z(t) \hat{J}_x(0) \rangle - \langle \hat{J}_x(t) \hat{J}_z(t) \rangle \langle \hat{J}_x(0) \rangle 
    &\simeq \langle \hat{J}_z(t) \rangle C_{J_x J_x}(t) + \langle \hat{J}_x(t) \rangle C_{J_z J_x}(t).
  \end{aligned}
\end{equation}
Then, inserting these back into Eq.~(\ref{app_eq:diss_ss_eom_1}), we obtain 
\begin{equation}
  \begin{aligned}
    \frac{1}{\Gamma }\partial_t C_{J_x J_x}(t) 
    &\simeq  - \frac{1}{2} \left[ (1 + \tanh(r))^2 - 2 (1 - \tanh(r)^2) \langle \hat{J}_z(t) \rangle \right] C_{J_x J_x}(t) + (1 - \tanh(r)^2)  \langle \hat{J}_x(t) \rangle C_{J_z J_x}(t)
  \end{aligned}
\end{equation}
Making the stationary approximation, we have $\langle \hat{J}_x(t) \rangle = \langle \hat{J}_x \rangle = 0$ \cite{Groszkowski_PRX_2022_dissipative_spin_squeezing, Pocklington_PRL_2025_adaptive_prep}, so 
\begin{equation}
  \begin{aligned}
    \frac{1}{\Gamma }\partial_t C_{J_x J_x}(t) 
    &\simeq  - \frac{1}{2} \left[ (1 + \tanh(r))^2 - 2 (1 - \tanh(r)^2) \langle \hat{J}_z(t) \rangle \right] C_{J_x J_x} \\
    &= \exp \left( - 2  \Gamma t \left[ 1 - 2 e^{-2r} \langle \hat{J}_z(t) \rangle^{(0)}  + O(e^{-4r})\right] \right) C_{J_x J_x}(0),
    &\simeq \exp \left( - 2  \Gamma t \right) C_{J_x J_x}(0),
  \end{aligned}
\end{equation}
where in the last line we have made use of the large $r$ approximation. Note that in the large $r$ regime, $\langle \Delta \hat{J}_x \rangle \sim J(J+1)/2$ \cite{Pocklington_PRL_2025_adaptive_prep}.

\subsection{Comments on the odd N}

In the main text, as well as the above, we have restricted ourselves to even $N$. We now provide a few comments on the case of odd $N$. 

For odd $N$, following \cite{Groszkowski_PRX_2022_dissipative_spin_squeezing, Agarwal_PRA_1990_OG_diss_ss}, the steady state may be obtained as follows. Let $\hat{L} = \hat{J}_+ - \tanh(r) \hat{J}_-$; then \begin{equation}\label{app_eq:odd_initial_state}
    \hat{\rho}_0 = \frac{(\hat{L}^{\dag} \hat{L})^{-1}}{{\rm Tr} ((\hat{L}^{\dag} \hat{L})^{-1})},
\end{equation}
is the steady state of the system. This state is neither a pure state nor a maximally mixed state, hence we are unable to use write down $I_{\rm E}(T)$ in a simple form like before. This suggests that odd $N$ is much less useful for sensing, as we lose the advantage of being able to perform direct photodetection on the output field to saturate the QFI. 

Nevertheless, preliminary numerical simulation of the $N$ odd case, with the same same parameters we used for the optimal sensing of the even $N$ case ($\Gamma = 1.89/2T$, $r = \ln(8N)$, but with initial state given by Eq.~(\ref{app_eq:odd_initial_state})) suggests that one also obtains an $N^2$ scaling for $I_{\rm E}(T)$ in the finite time regime, albeit with a smaller prefactor than the even $N$ case. This is surprising, since the steady state does not contain any spin squeezing \cite{Groszkowski_PRX_2022_dissipative_spin_squeezing}. While the $N^2$ scaling in the $N$ even case can be attributed to spin squeezing in the steady state, any $N^2$ scaling in the odd $N$ case should be attributed to entanglement with the environment, similar to the high-temperature sensor. We note also that one may be able to pick a smaller value of $r$ that allows us to avoid the odd-even effect while maintaining an $N^2$ scaling, as is done in \cite{Groszkowski_PRX_2022_dissipative_spin_squeezing}.

\subsection{Photodetection is optimal for system with dark states}\label{app_sec:photodetection}

We note that the proof of optimality of photodetection on the cascaded setup in Ref.~\cite{Yang_PRX_2023_CQA_sensing} requires no modification to show that a direct photodetection on a system with a dark state is optimal. Here, for completeness, we provide a proof of the same fact from the point of view of the non-Hermitian Hamiltonian.

Our strategy will be to explicitly calculate the classical Fisher information associated with the probability distribution of either detecting one or more photons ($P_1$), or detecting zero photons ($P_0$). This probability distribution is obtained from initializing the system in the pure state $| \Psi \rangle$, which is a dark state of the effective non-Hermitian Hamiltonian $\hat{H}_{\rm eff}$, and then subjecting it to evolution via $\theta \hat{Z}$ some time interval $T$. We will show that this classical Fisher information matches the general expression for the dark state QFI obtained in Eq.~(\ref{app_eq:intermediate_I_E_ss}).

Before we look at the state, we note that for a binary distribution where the probability has the functional form \begin{equation}
  P_0(\theta) = 1 + C_2 \theta^2 + O(\theta^3),
\end{equation}
we have the classical Fisher information \begin{equation}\label{eq:classical_binary_FI}
  I_{\rm Cl} = -4 C_2,
\end{equation}
where $C_2 < 0$. This can be obtained from the definition of the classical Fisher information.

The probability of detecting no photon, i.e. $P_0$, can be obtained by calculating the norm of the state evolving under $\hat{H}_{\rm eff} + \theta \hat{Z}$ for a time $T$, namely,
\begin{equation}
    | \Psi(T) \rangle = e^{- i \hat{H}_{\rm eff} T - i \theta \hat{Z} T} | \Psi \rangle,
\end{equation}
where we recall that $\hat{H}_{\rm eff}$ is defined in terms of the Hamiltonian $\hat{H}_0$ and the jump operators $\{\hat{L}_i\}$ of the system as 
\begin{equation}
    \hat{H}_{\rm eff} = \hat{H}_0 - \frac{1}{2} i \sum_{i} \hat{L}^{\dag}_i \hat{L}_i
\end{equation}
Now, to $O(\theta^3)$, this is simply the (conjugate of the) bra that appears in Eq.~(\ref{app_eq:dyson_H_eff}); recall this is obtained via a Dyson expansion of the propagator $e^{- i \hat{H}_{\rm eff} T - i \theta \hat{Z}}$, 
\begin{equation}
    | \Psi(T) \rangle = 
    \left( 1 - i \theta \int_0^T dt_1 \hat{Z} e^{- i \hat{H}_{\rm eff} t_1} - \theta^2 \int_0^T dt_1 \int_0^{t_1} dt_2 \hat{Z} e^{-i \hat{H}_{\rm eff} (t_1 - t_2)} \hat{Z} e^{-i \hat{H}_{\rm eff} t_1} \right) | \Psi \rangle + O(\theta^2).
\end{equation}
Also from Eq.~(\ref{app_eq:dyson_H_eff}) we then see that we can write 
\begin{equation}
    | \Psi(T) \rangle = \langle \Psi | \mu(T) + O(\theta^3).
\end{equation}

The norm of this state gives $P_0$, i.e. 
\begin{equation}
    P_0 = \langle \Psi(T) | \Psi(T) \rangle 
    =\langle \Psi | e^{ i \hat{H}_{\rm eff}^{\dag} T + i \theta \hat{Z}} e^{- i \hat{H}_{\rm eff} T - i \theta \hat{Z}} | \Psi \rangle = {\rm Tr} (\hat{\mu}(T) \hat{\mu}(T)^{\dag}) + O(\theta^3).
\end{equation}
The rest is almost immediate. From Eq.~(\ref{app_eq:full_ptb_1b}), we recognize this as 
\begin{equation}
\begin{aligned}
        P_0 = &1 + \theta^2  \int_0^T d t_1 \int_0^T dt_2 \left\langle \hat{Z} e^{i \hat{H}_{\rm eff}^{\dag} t_1} e^{- i \hat{H}_{\rm eff} t_2} \hat{Z} \right\rangle   \\
        &- \theta^2  \int_0^T d t_1 \int_0^{t_1} dt_2 C_{ZZ}(t_1 - t_2) + O(\theta^3),
\end{aligned}
\end{equation}
and from Eq.~(\ref{eq:classical_binary_FI}), we read off
\begin{equation}
    I_{\rm Cl}(T) = 4  \int_0^T d t_1 \int_0^{t_1} dt_2 C_{ZZ}(t_1 - t_2)  - 4 \int_0^T d t_1 \int_0^T dt_2 \left\langle \hat{Z} e^{i \hat{H}_{\rm eff}^{\dag} t_1} e^{- i \hat{H}_{\rm eff} t_2} \hat{Z} \right\rangle^{(0)} + O(\theta),
\end{equation}
exactly equal (to $O(\theta)$) to the expression Eq.~(\ref{app_eq:intermediate_I_E_ss}) we derived for the $I_{\rm E}$ of a system hosting a dark state. Hence when $\theta \rightarrow 0$, we have that photodetection is optimal for dark states.



\end{document}